\documentclass{ieeeaccess}
\usepackage{cite}
\usepackage{listings}
\usepackage{xcolor}
\usepackage{caption}
\usepackage{float}
\usepackage{amsmath,amssymb,amsfonts}
\usepackage{algorithmic}
\usepackage{graphicx}
\usepackage{textcomp}
\usepackage[none]{hyphenat}
\usepackage{colortbl}

\def\BibTeX{{\rm B\kern-.05em{\sc i\kern-.025em b}\kern-.08em
    T\kern-.1667em\lower.7ex\hbox{E}\kern-.125emX}}

\usepackage{subcaption}
\usepackage{makecell}
\usepackage{hhline}
\usepackage[
    colorlinks=true,            % Ativa cores nos links
    linkcolor=blue,             % Cor para links internos (ex.: sumário)
    citecolor=blue,             % Cor para citações
    filecolor=magenta,          % Cor para links de arquivos
    urlcolor=black, %cyan,      % Cor para URLs
    pdfpagemode=UseOutlines     % Define o modo de abertura (ex.: bookmarks abertos)
]{hyperref}

\begin{document}
\history{Date of publication xxxx 00, 0000, date of current version xxxx 00, 0000.}
\doi{10.1109/ACCESS.2025.DOI}

\title{Objective Video Quality Assessment in FWA-based Over-the-Top Content Delivery across Open Source 5G Networks}

\author{\uppercase{Nelson Ion}\authorrefmark{1,2}, \uppercase{Flávio Silva}\authorrefmark{3}, \uppercase{Paulo Silva}\authorrefmark{4}, \uppercase{Ricardo Silva}\authorrefmark{4}, \uppercase{Mathews Lima}\authorrefmark{1}, \uppercase{Daniel Luna}\authorrefmark{1},\uppercase{Antonio Campos}\authorrefmark{3,4}, \uppercase{Augusto Neto\authorrefmark{1}, and Vicente Sousa}\authorrefmark{3,4}.}
\address[1]{Graduate Program in Systems and Computing (PPgSC), Federal University of Rio Grande do Norte, Natal, 59078-970, Rio Grande do Norte, Brazil.}
\address[2]{Digital Metropolis Institute (IMD), Federal University of Rio Grande do Norte, Natal, 59078-970, Rio Grande do Norte, Brazil.}
\address[3]{Department of Communications Engineering, Federal University of Rio Grande do Norte, Natal, 59078-970, Rio Grande do Norte, Brazil.}
\address[4]{Graduate Program in Electrical and Computer Engineering (PPgEEC), Federal University of Rio Grande do Norte, Natal, 59078-970, Rio Grande do Norte, Brazil.}

\tfootnote{This research was carried out under the LANCE (Leading Advanced Technologies Center of Excellence) facilities, and partially funded by the \textit{Coordenação de Aperfeiçoamento de Pessoal de Nível Superior - Brasil} (CAPES) - Finance Code 001. Moreover, this research is part of the INCT of Intelligent Communications Networks and the Internet of Things (ICoNIoT) funded by CNPq (proc. 405940/2022-0), CAPES Brasil, Finance Code 88887.954253/2024-00, and by CNPq grant \#405940/2022-0. In compliance with CNPq Ordinance No. 2664/2026, the authors attest that generative artificial intelligence was used solely for linguistic editing, text refinement, and the generation of preliminary figure drafts, with no usage in the study’s conceptualization, analysis, or scientific interpretation. The Article Processing Charge (APC) for the publication of this research was funded by the Coordenação de Aperfeiçoamento de Pessoal de Nível Superior - CAPES (ROR ID: 00x0ma614).}

\markboth
{Ion \headeretal: Objective Video Quality Assessment in FWA-based Over-the-Top Content Delivery across Open Source 5G Networks}
{Ion \headeretal: Objective Video Quality Assessment in FWA-based Over-the-Top Content Delivery across Open Source 5G Networks}

\corresp{Corresponding author: Vicente Sousa (e-mail: vicente.sousa@ufrn.br).}

\begin{abstract}

This paper presents a comprehensive experimental video quality assessment in 5G-based Fixed Wireless Access (FWA) networks, leveraging a real open source 5G network testbed. Addressing the critical need for robust UHD video delivery, the study systematically investigates the impact of transmission modes (SISO and MIMO), modulation schemes (64-QAM and 256-QAM), and network load (one to five users) on both network throughput and perceived video quality. Unlike simulation-based approaches, our methodology integrates automated streaming campaigns with objective full-reference Quality of Experience (QoE) metrics (PSNR, SSIM, and VMAF) computed from 4K video streams delivered via MPEG-DASH, enabling frame-level correlation between network performance and user experience. Key findings demonstrate that MIMO consistently outperforms SISO, sustaining higher throughput and superior visual quality even under increasing user demand. Specifically, MIMO with 256-QAM maintained VMAF scores above the visually lossless threshold with five concurrent users, while SISO configurations experienced significant degradation. These results provide empirical evidence for the critical role of spatial diversity and higher-order modulation in ensuring reliable UHD video delivery over FWA. The proposed reproducible framework for detailed quality assessment offers valuable insights for future optimization of 5G FWA deployments, contributing to the practical understanding and deployment strategies for high-quality video services in Open RAN environments.
\end{abstract}


\titlepgskip=-15pt

\maketitle

\section{Introduction}

Internet-based Over-the-Top (OTT) applications have considerably increased the density of multimedia flows in recent years, primarily due to advancements in compression technology and the emergence of more efficient and cost-effective transmission systems. This accounts for a significant portion of approximately 65\% of all Internet traffic~\cite{sandvine2023}. The growing demand for high-quality OTT services has intensified the need for advanced networking solutions, capable of sustaining stringent Quality of Service (QoS) requirements to ensure an acceptable Quality of Experience (QoE) for end-users. Throughout this manuscript, the term Ultra High Definition (UHD) refers specifically to multimedia flows featuring a resolution of 3,840 × 2,160 pixels, which we also denote as 4K to maintain alignment with both academic and industry terminology. In order to ensure a consistent and acceptable QoE for these UHD flows, the network must accommodate an intense data rate of at least 25 Mbps~\cite{navarro2020survey}. 

While optical fiber infrastructures remain the gold standard for high-capacity connectivity, their deployment is often limited by high costs, complex installation processes, and maintenance challenges. As a result, alternative technologies that can deliver comparable performance with greater flexibility and lower costs have gained relevance. Among these, 5G mobile networking stands out for its enhanced capabilities in bandwidth, latency, reliability, and scalability—features that align closely with the stringent needs of modern OTT applications.

\begin{figure}[ht]
\centering
\includegraphics[width=\linewidth]{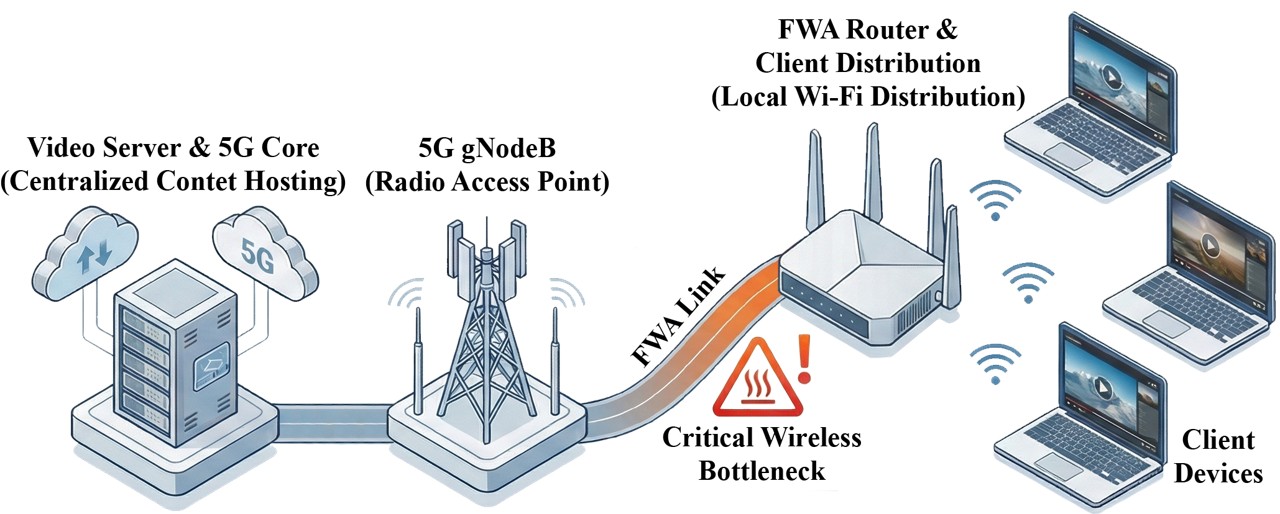}
\caption{Conceptual FWA deployment model (figure produced with assistance from an OpenAI generative tool).}
\label{fig:fwa}
\end{figure}

Within this context, Fixed Wireless Access (FWA) has emerged as a promising and economically viable solution for extending broadband connectivity. Leveraging the advancements of 5G New Radio (5G-NR) technology, as illustrated in Figure~\ref{fig:fwa}, FWA provides high-speed wireless links between Internet Service Providers (ISPs) and end users, particularly in regions where wired infrastructure remains unavailable or impractical~\cite{topyan2020architectural}. Its combination of mobility, rapid deployment, and high throughput positions FWA as a key enabler for flexible and high-quality OTT content delivery~\cite{gsma2023ts64}. By bridging the gap between performance and accessibility, FWA supports the widespread distribution of OTT UHD data-intensive services, expanding the reach of next-generation digital entertainment and communication ecosystems~\cite{gsma2023ts64}. 

Despite the growing potential of FWA broadband mobile connectivity, the delivery of OTT UHD content over such links still faces several challenges~\cite{survey_2021}. FWA networks operate in wireless environments that are inherently variable, where channel conditions fluctuate due to interference, user mobility, weather effects, and signal attenuation. When bandwidth is shared among multiple users, this variability can cause congestion and throughput degradation, making it difficult to sustain the high and stable data rates required for UHD streaming. 

Moreover, FWA systems must manage contention between diverse traffic types (e.g., video, voice, and IoT) while maintaining low latency and minimizing jitter, both of which are essential for continuous playback and consistent QoE. The dynamic nature of wireless channels also jeopardizes resource allocation and scheduling, requiring adaptive mechanisms capable of responding to rapid variations in link quality. These challenges highlight the importance of intelligent traffic management, adaptive bitrate streaming, and quality assessment mechanisms to ensure reliable and high-quality OTT service delivery over FWA infrastructures.

Thus, despite the growing body of research on multimedia delivery, existing QoE studies frequently operate under three dominant assumptions that limit their practical applicability to FWA context: i) a heavy reliance on simulation-based or emulator-centric evaluations, which often bypass the complex impairments of a real wireless RF-chain; ii) a QoS-centric focus that prioritizes high-level indicators like delay and jitter, leaving a persistent gap in understanding fine-grained, frame-level perceived quality; and iii) the use of controlled-sample scenarios that do not reflect the contention of shared-link environments and congestion inherent in multi-user FWA environments. These limitations become particularly critical in 5G FWA scenarios supporting UHD OTT services, as 4K content requires an intense data rate. In a live wireless environment, the channel's dynamism and unpredictability make the 5G link a critical bottleneck, and simulations often fail to capture how these physical-layer realities interact with hardware-specific constraints. Consequently, there is an urgent need for empirical evidence to quantify how spatial diversity and modulation schemes act as the quality ceiling for multiple concurrent UHD users.

In this paper, we start from the premise that ensuring network QoS to support OTT UHD content delivery can be enhanced through the adoption of reliable Video Quality Assessment (VQA) methods~\cite{video_quality_2023}. VQA tools allow OTT service providers to gain deeper insights into content delivery performance and identify potential impairments that may affect the end-user experience. By systematically monitoring and improving video quality, providers can deliver a more consistent and satisfying QoE, ultimately fostering stronger user engagement and profitability.

Given the growing consumption of OTT content on mobile devices, this study presents a comprehensive experimental evaluation of video quality in 5G-based FWA networks using a real 5G telco-cloud testbed. The primary objective is to systematically investigate how different transmission modes (SISO and MIMO), modulation schemes (64-QAM and 256-QAM), and network load conditions impact both network throughput and the perceived QoE of OTT UHD content delivery in FWA environments.

The contributions of this paper are threefold. First, we advance beyond traditional simulation-based analyses by conducting an empirical evaluation on a real 5G telco-cloud testbed, promoting practical insights into FWA performance under realistic conditions. Second, we propose a novel methodology that integrates automated streaming campaigns with objective, full-reference QoE metrics, namely Peak Signal-to-Noise Ratio (PSNR), Structural Similarity Index (SSIM) and Video Multi-method Assessment Fusion (VMAF), for OTT UHD content (4k video streams) delivered via MPEG-Dynamic Adaptive Streaming over HTTP (DASH). This framework enables fine-grained, frame-level correlation between network performance and perceived quality, providing a deeper understanding than conventional QoS indicators. Finally, we present empirical evidence demonstrating the consistent performance advantage of MIMO over SISO, and highlight the critical role of 256-QAM modulation in maintaining a high QoE of OTT UHD content under increasing user demand, achieving visually lossless VMAF scores even with multiple concurrent users.

The remainder of this paper is organized as follows. Section~\ref{sec:rela_works} provides an analysis of related work. Section~\ref{sec:evaluation-and-metrics} details the evaluation metrics and scenarios. Section~\ref{sec:methodology} describes the experimental methodology. Section~\ref{sec:results} presents a discussion on the experimental outcomes. Finally, Section~\ref{sec:conclusions} concludes the paper and outlines future work.

\section{Related Work}
\label{sec:rela_works}

This section presents a review of the literature in research endeavors on the field of QoE assessment over UHD OTT content delivery, aiming to position the present study within the context of related research efforts.

The authors in~\cite{khan2021ultra} investigated UHD video streaming over 5G last-mile networks integrated with FWA systems, focusing on link performance under congested network conditions and various adaptive bitrate (ABR) protocols. Conducted in a simulated environment, the study identified the Buffer Occupancy-based Lyapunov Algorithm (BOLA) as the most effective ABR scheme for maintaining user-perceived quality based on subjective QoE evaluation. However, the reliance on simulations limits the applicability of the results to real-world operator environments, where wireless variability, interference, and user mobility introduce complexities that are difficult to emulate accurately.

A complementary contribution is presented in~\cite{moina2023video}, which introduces the Video Quality Metrics Toolkit (VQMTK)\footnote{https://github.com/cloudmedialab-uv/vqmtk}, a platform for objective QoE assessment that integrates up to 14 video quality metrics within a unified interface. The toolkit’s capability to automate metric computation through web and command-line modes makes it particularly suitable for reproducible research. While it offers methodological advances, its validation was restricted to controlled 4K video samples, not extending to real 5G or FWA network deployments.

In~\cite{leszczuk2024objective}, authors propose a classification framework for objective VQA methods, encompassing full-reference and reduced-reference models. It categorizes methods according to their consideration of natural scene statistics or perceptual features aligned with the human visual system (HVS) and provides an extensive comparative analysis across video resolutions. This theoretical foundation supports the understanding of objective quality assessment but lacks experimental validation under mobile or wireless conditions.

In~\cite{cross-relation_2020}, the authors examined the cross-correlation between QoS indicators (e.g., delay, jitter, and packet loss) and their impact on QoE in video streaming. Using an extended version of the NetEm emulator, they modeled dependencies between delay and packet loss, demonstrating that even small variations (0.05\%–1.5\%) can significantly degrade perceived quality. While insightful, such emulator-based studies still fail to capture the full dynamism and unpredictability of live 5G network behavior.

Beyond the physical and link-layer impacts addressed in this study, the broader ecosystem of QoE enhancement includes higher-level optimization strategies such as network-aware path control and edge delivery mechanisms. For instance, the authors of \cite{sdn_routing_2023} have proposed SDN-based routing adaptation frameworks that dynamically optimize network paths to enhance the QoE of multimedia streaming services. While such network-layer path optimizations are vital, they operate on the assumption of reliable underlying link performance. Similarly, the implementation of redirection mechanisms, such as FASP/HTTP, and Modified Cuckoo Search protocol (MCSA) on CDN edge servers plays a critical role in stabilizing adaptive video streaming by connecting users to the ideal server more quickly~\cite{cdn_edge_2023}. Our work complements these infrastructure-level delivery strategies by providing an empirical analysis of how fundamental 5G-NR physical-layer configurations, specifically MIMO and 256-QAM, dictate the performance boundaries of the 5G FWA link, providing the necessary grounding for these higher-level network and edge optimizations in real-world deployments.

Collectively, these studies reinforce the relevance of analyzing the QoS–QoE relationship in UHD video streaming but reveal a persistent gap between simulation-based research and operational network realities.
While many previous works rely on simulation or emulation to study ABR protocols and QoS/QoE correlations, while valuable for controlled experimentation, such approaches often fall short of providing the actionable insights into physical and link-layer performance that network operators require in real-world deployments. To bridge this gap, our work shifts the focus toward evaluating the empirical impacts of the physical and link layers—specifically the interplay between MIMO versus SISO configurations and modulation orders—within a real 5G environment, advancing the state of the art by conducting a real-world experimental assessment of UHD OTT video delivery over 5G FWA networks. 
By leveraging a live 5G telco-cloud open-source testbed, this study provides empirical evidence of how network parameters influence objective VQA models, offering a more realistic perspective on QoE behavior under practical deployment scenarios. Table~\ref{tab:comparison_related_work} provides a comparison of related works and the investigation herein proposed.

\begin{table}[h!]
\centering
\caption{Comparison of related research.}
\label{tab:comparison_related_work}
\begin{tabular}{c|c|c|c}
\hline
\textbf{Reference} & \textbf{Environment} & \textbf{\makecell{Evaluation \\ Focus}} & \textbf{Architecture} \\ \hline
\cite{khan2021ultra} & Simulation & \makecell{Subjective \\ QoE} & RAN Emulated \\ 
\cite{moina2023video} & Dataset & \makecell{Objective \\ QoE} & N/A \\
\cite{leszczuk2024objective} & Dataset & \makecell{VQA \\ Framework} & N/A \\ 
\cite{cross-relation_2020} & Emulator & \makecell{QoS vs QoE} & Full Emulated \\ 
\textbf{This Work} & \textbf{Real Testbed} & \textbf{QoS vs. QoE} & \makecell{\textbf{ 5G Open RAN} \\ \textbf{with FWA}} \\ \hline
\end{tabular}
\end{table}

\section{Outlook in QoE Assessment Metrics and Scenarios}
\label{sec:evaluation-and-metrics}

Serving as the theoretical basis of our methodology, this section outlines the evaluation metrics and scenarios underpinning the video assessment in this paper.

VQA can be performed using subjective or objective methods~\cite{min2024perceptual}. Subjective assessment, which relies on human observers, remains the most reliable and accurate means of evaluating perceived quality. However, it is time-consuming, resource-intensive, and often impractical for large-scale or real-time applications. As a result, objective assessment methods, which automatically estimate perceived video quality of OTT content, have gained prominence in both research and industry settings~\cite{min2024perceptual, chikkerur2011objective, leszczuk2024objective, vranjes2007objective, Aguilar2024}.

Video quality is influenced by a wide range of factors, including spatial and temporal resolution, frame rate, blur, noise, and compression efficiency. In addition, the type of OTT content, such as user-generated videos, virtual or augmented reality (VR/AR), high frame rate (HFR) media, or audiovisual productions, plays a significant role in determining perceived quality. The HVS responds differently to distortions depending on these characteristics, resulting in high variability in how degradations affect perception across content categories. This variability makes it challenging to design objective VQA models that generalize well across scenarios~\cite{min2024perceptual, chikkerur2011objective, leszczuk2024objective, vranjes2007objective, Aguilar2024}.

A large body of research has examined subjective and objective approaches under diverse conditions and influencing factors~\cite{min2024perceptual, chikkerur2011objective, leszczuk2024objective, vranjes2007objective, Aguilar2024}. These studies have been instrumental in advancing OTT applications across compression, transmission, display technology, content analysis, and recording systems. Within this landscape, objective full-reference metrics have become the most widely adopted tools for assessing video quality and distortion, as they offer reproducibility and ease of implementation.

Among objective metrics, the Mean Squared Error (MSE) and the PSNR are the most common due to their mathematical simplicity, clear physical meaning, and compatibility with optimization algorithms. MSE measures the average squared difference between corresponding pixels of the original and processed frames, with lower values indicating less distortion and thus better quality~\cite{vranjes2007objective, leszczuk2024objective, wang2003multiscal}. PSNR, derived from MSE, expresses the ratio between the maximum possible signal power and the noise power introduced during processing or transmission. It is expressed in decibels (dB), where higher PSNR values correspond to better visual quality~\cite{vranjes2007objective, leszczuk2024objective, wang2003multiscal}. The PSNR is computed as follows~\cite{Rassool2017}:

\begin{equation}
\text{PSNR} = 20 \log_{10}\left( \frac{\text{Max}_I}{\text{MSE}} \right)
\end{equation}

\noindent where Max$_{I}$ is the maximum possible pixel value of the image (typically 255 for 8-bit images). 

PSNR values above 30 dB generally correspond to good visual quality, whereas values below 20 dB typically indicate noticeable degradation. Despite its simplicity and widespread adoption, PSNR often exhibits a weak correlation with human visual perception, as it does not account for the structural and perceptual characteristics of the content.

To achieve a more accurate assessment of video quality, full-reference perceptual metrics closely-correlated with human visual perception, such as the SSIM and VMAF~\cite{wang2004image, wang2004video, wang2002video}. These metrics compare a degraded or compressed video against its original version, providing a perceptually aligned estimation of quality loss. Unlike traditional distortion-based measures such as MSE and PSNR, SSIM and VMAF incorporate HVS principles, resulting in a stronger correlation with perceived QoE~\cite{wang2004image, wang2004video, wang2002video}.

The SSIM metric, in particular, evaluates image or video quality by jointly considering luminance, contrast, and structural similarity between reference and processed frames. This multi-dimensional comparison enables SSIM to more accurately reflect the way humans perceive visual similarity, making it a preferred choice in perceptual video research and quality assessment. To be considered reliable, the SSIM index must satisfy the following fundamental conditions~\cite{leszczuk2024objective}:

\begin{itemize}
  \item $\text{SSIM}(x, y) = \text{SSIM}(y, x)$;
  \item $\text{SSIM}(x, y) \leq 1$;
  \item $\text{SSIM}(x, y) = 1$ if and only if $x = y$.
\end{itemize}

When one of the compared images is assumed to represent a reference of perfect quality, the SSIM index provides a quantitative measure of the relative quality of the other image. Its values range from –1 to 1, where 1 denotes perfect structural similarity~\cite{leszczuk2024objective}. In practice, SSIM values between 0.95 and 1.00 correspond to excellent visual quality, where the reconstructed or processed video is virtually indistinguishable from the original. Values in the range of 0.90 to 0.95 indicate very good quality, with only minor perceptual differences that are barely noticeable to viewers. Scores between 0.80 and 0.90 typically represent good quality, where slight artifacts may appear, but the overall viewing experience remains acceptable for most users.

To achieve an even stronger alignment with human perceptual judgment on the OTT content, the VMAF metric employs machine learning techniques to fuse multiple quality indicators into a single perceptual score. Developed and widely adopted by the industry, most notably by Netflix and YouTube, VMAF combines features such as detail loss, blockiness, and temporal smoothness into a unified score ranging from 0 to 100~\cite{wang2004video}, where:

\begin{itemize}
  \item Scores between 90 and 100 indicate \textbf{excellent} quality.
  \item Scores between 70 and 90 indicate \textbf{good} quality.
  \item Scores below 70 indicate \textbf{noticeable degradation}.
\end{itemize}

To better understand the performance boundaries of the 5G FWA link for UHD OTT service delivery, this study investigates how the system adapts under varying capacity and network load conditions. In particular, we examined the effects of different 5G multi-antenna transmission modes and modulation settings, as well as the impact of scaling the number of simultaneously connected users. Using distinct parameter combinations, twelve experimental campaigns were conducted, each featuring ten independent execution runs to ensure statistical robustness. The complete set of parameter configurations used in these experiments is summarized in Table~\ref{tab:campanhas}. This systematic approach allows for a granular investigation into how spatial diversity (SISO vs. MIMO with two transmitting and two receiving antennas), higher-order modulation (64-QAM vs. 256-QAM), and scaling network load (emulating typical household or small-office contention for one to five users) interact to define the 5G physical-layer quality ceiling required to sustain the bitrates necessary for 4K delivery, directly quantifying the impact of these parameters on both network-level QoS and perceived QoE in operational FWA infrastructures. 

\begin{table}[!h]
\small
\caption{Parameters variation of the campaigns.}
\label{tab:campanhas}
\centering
\begin{tabular}{ c|c }
 \hline
 \textbf{Parameter} & \textbf{Variations}\\
 \hline
 Transmission Mode & SISO or 2$\times$2 MIMO\\
 Modulation of Max. MCS & 64-QAM or 256-QAM  \\
 Number of Users & 1, 3 or 5\\
 \hline
\end{tabular}
\end{table}

To ensure that perceptual conclusions such as \textit{visually lossless} are technically grounded, we align our objective metric results with established HVS benchmarks~\cite{min2024perceptual, chikkerur2011objective, leszczuk2024objective, vranjes2007objective, Aguilar2024}. While subjective user studies are the most accurate means of evaluating perceived quality, they are often impractical for the high-granularity, frame-level analysis required in large-scale experimental campaigns. Therefore, we define visually lossless quality as achieving a VMAF score of 93\% or higher, a threshold recognized in the literature and industry as being visually indistinguishable from the reference video. This is further supported by the SSIM index, where values between 0.95 and 1.00 are categorized as excellent quality, representing a reconstructed video that is virtually indistinguishable from the original for most viewers. By linking our findings to these specific numerical thresholds, we provide a validated perceptual context for the objective results obtained across different 5G FWA scenarios.

\section{Testbed Experimental Methodology}
\label{sec:methodology}
\subsection{Setup Configuration} 

\begin{figure*}[ht]
\centering
\includegraphics[width=\linewidth]{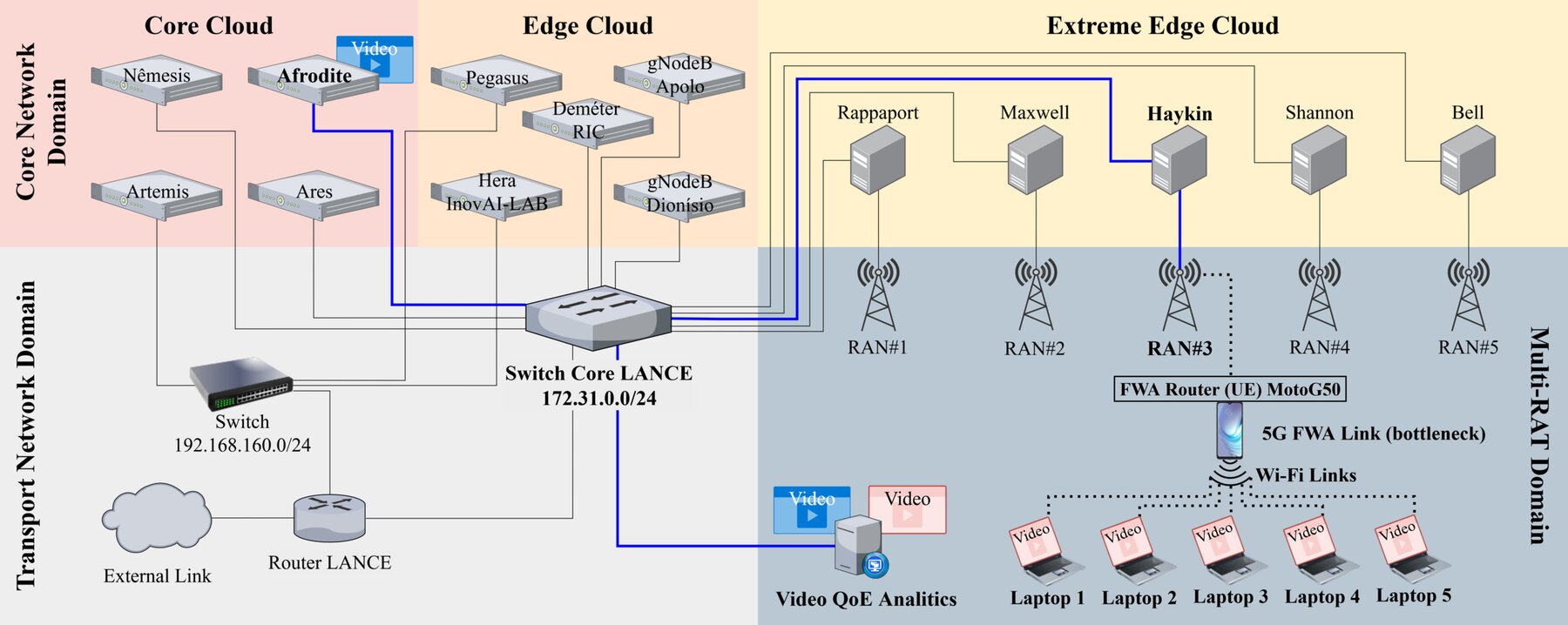}
\caption{LANCE's 5G testbed for 4K video quality assessment.}
\label{fig:testebed}
\end{figure*}

The assessment of OTT application's video quality metrics presented in this paper was performed using a real 5G telco-cloud open-source testbed at the Leading Advanced Technologies Center of Excellence (LANCE)\footnote{lance.ufrn.br}, which is illustrated in Figure~\ref{fig:testebed}. 

The experimental deployment was introduced on the node named Haykin, which entails the extreme edge domain of LANCE's telco-cloud testbed. This node lies on a Desktop equipped with an AMD Ryzen 7 5700G CPU, 128~GB of RAM, a Kingston SNV2S2000G NVMe drive, and an Ubuntu Server~22.04 LTS operating system. The entire 5G system, consisting of a core network (5GC) implemented with Open5GS~\cite{open5gs2025} and a 5G radio access network (gNB) based on the srsRAN protocol stack~\cite{srsran2025}, is deployed at the Haykin node of the extreme edge testbed premises. The radio frequency (RF) front-end employs a Universal Software Radio Peripheral (USRP) Ettus B210, interfaced via the USRP Hardware Driver API. A Motorola G50 5G smartphone emulates the FWA router capability, which receives the 5G signal from the gNB and forwards traffic over Wi-Fi to client stations. These client stations consist of laptops equipped with Wi-Fi~5 (IEEE 802.11ac) interfaces. Following the principles of cloud-based architectures, the video server is hosted at the Afrodite server node located in the core cloud domain of LANCE's telco-cloud testbed to leverage centralized high-scale processing, scalable storage, and simplified management of OTT video backend server. The Afrodite server runs the Ubuntu Server~22.04 LTS operating system and features an Intel(R) Xeon(R) Platinum 8461V processor, 128~GB of RAM, and a Samsung MZILG3T8HCLS SAS drive with a sequential read speed of up to 4200 MB/s.

In what concern the characteristics of the OTT application, the input video is encoded in five resolutions (4K, FullHD, HD, SD, and nSD lower resolution), each associated with a specific target bitrate, namely 4k (14 and 18~Mbps), FullHD (10 and 12~Mbps), HD (6 and 8 Mbps), SD (3, 4 and 5~Mbps), and nSD (1.8, 2 and 2.5~Mbps). The audio, in turn, is encoded using Advanced Audio Coding (AAC) at 128 kbps, and the video segments are packaged following the MPEG-DASH standard with a 2-second segment duration. All experimental scripts were executed in a Linux-based environment equipped with FFmpeg, FFprobe, and MP4Box tools for video processing metric extraction. To ensure reproducibility and systematic evaluation, we defined an experimental methodology composed of eight sequential steps, which are summarized as follows:

\begin{enumerate}
    \item Configuration of the MPEG-DASH server environment.
    \item Audio encoding.
    \item Video encoding.
    \item DASH manifest generation.
    \item Packet capture.
    \item Video reconstruction.
    \item Video metric computation.
    \item Network metric acquisition.
\end{enumerate}

\subsection{MPEG-DASH Server Configuration}

The DASH server operates using \textit{Nginx}, which is running inside a Docker container hosted on the Afrodite server at the Core Cloud. This server delivers MPEG-DASH segments generated from a 3-minute video (10,800 frames), previously encoded with \textit{FFmpeg}.

The process of generating adaptive video representations uses the script \texttt{generate\allowbreak\_dash\_manifest.sh}~\cite{ufrn2025dash}, which executes the following steps: (i) extracts and encodes the audio track in the \texttt{AAC} format at 128~kbps; (ii) creates multiple video representations at 2160p, 1080p, 720p, 480p, and 360p resolutions, with bitrates ranging from 1.8~Mbps and 18~Mbps; and (iii) segments the encoded files with \texttt{MP4Box} using the \texttt{dashavc264:live} profile, producing the manifest file (\texttt{.mpd}) and media segments (\texttt{.m4s}) required for adaptive streaming via MPEG-DASH.

The experimental environment uses the \texttt{docker-compose.yml}~\cite{ufrn2025dash} configuration file to instantiate the \texttt{nginx:alpine} container with dedicated volumes for: (i) the static DASH media files; (ii) the HTML player \texttt{index.html}; (iii) the \textit{Nginx} configuration file (\texttt{nginx.conf}); and (iv) the customized \texttt{mime.types} definitions. The \texttt{nginx\allowbreak.conf} file includes the directive \verb|add_header Cache-Control “no-cache”| to disable caching and enable reproducible performance testing under different network conditions. The \texttt{mime.types} file defines the correct MIME associations for \texttt{.mpd}, \texttt{.m4s}, and \texttt{.mp4} files, which are essential for browser-based playback.

The HTML player uses the \texttt{dash.js} library and supports dynamic switching between ABR algorithms such as \texttt{BOLA} and \texttt{Oboe}, enabling real-time experimental control. This experiment selects the BOLA algorithm as the representative ABR mechanism for this study because it is a state-of-the-art solution recognized for its stability and effectiveness in maintaining user-perceived quality within congested last-mile bottlenecks~\cite{khan2021ultra}.
However, the player design remains modular and extensible, thus allowing future comparative evaluations of additional adaptation strategies, including throughput-based algorithms.

\subsection{Audio and Video Encoding Settings}

The original video, obtained from the Blender Foundation repository~\cite{bbb3d2025video}, is encoded with \textit{FFmpeg} using parameters compatible with the \texttt{dashavc264:live} profile, in order to enable adaptive streaming via MPEG-DASH. The OTT video is trimmed to a duration of 3 minutes (corresponding to 10,800 frames at 60~fps) to standardize the test dataset and reduce the total processing time.

The audio track is extracted and encoded in \texttt{AAC} format at 128~kbps, ensuring compatibility with modern web browsers. The video is then transcoded into multiple representations at 2160p, 1080p, 720p, 480p, and 360p resolutions, with bitrates ranging from 1.8~Mbps to 18~Mbps, to address different network conditions and device capabilities. In the context of MPEG-DASH, each \textit{representation} corresponds to a distinct combination of resolution and bitrate, time-aligned to support adaptive switching during playback. Table~\ref{tab:bitrates} summarizes the resolutions and respective encoding rates adopted in this research for the OTT application service.

\begin{table}[!ht]
\small
\caption{Video representations: Resolutions and bit rates used in encoding.}
\label{tab:bitrates}
\centering
\begin{tabular}{c|c}
 \hline
 \multicolumn{1}{c|}{\textbf{Resolution}} & \multicolumn{1}{c}{\textbf{Bit Rate (Mbps)}} \\
 \hline
 3840$\times$2160 (4K) & 18, 16, 14 \\
 1920$\times$1080 (FHD) & 12, 10 \\
 1280$\times$720 (HD) & 8, 6 \\
 854$\times$480 (SD) & 5, 4, 3 \\
 640$\times$360 (Low) & 2{.}5, 2, 1{.}8 \\
 \hline
\end{tabular}
\end{table}

All the video segments use the \texttt{H.264/AVC} codec with the \texttt{fast} preset, a fixed \textit{GOP} size of 120 frames, and the parameter \texttt{sc\_threshold} = 0 to maintain uniform intervals between keyframes. These settings ensure synchronization among the multiple representations (video versions) and enable accurate segmentation with \texttt{MP4Box}. The chosen configuration maintains compatibility with modern MPEG-DASH players and allows controlled variations in visual quality during adaptive streaming of the OTT content.

The choice of the \textit{Big Buck Bunny} sequence was made to provide a standardized test dataset and ensure a computationally feasible frame-level analysis across the multiple experimental scenarios. However, we acknowledge that this represents a limitation, as different types of OTT content, such as high-motion sports, low-light scenes, or VR/AR, can significantly influence perceived quality. Because the HVS responds differently to distortions depending on these characteristics, the performance of objective VQA models can vary across content categories. Consequently, while our current results provide a robust baseline for animation, they may not fully capture the perceptual variability inherent in more dynamic or complex video scenes.

\subsection{DASH Manifest Generation}

The \texttt{MP4Box} tool generated the DASH manifest and packaged the encoded audio and video representations in a format compatible with the MPEG-DASH standard. The command used is implemented in the script \verb|generate_dash_manifest.sh|~\cite{ufrn2025dash} and includes the main parameters needed for temporal segmentation and construction of the manifest, as shown in Listing~\ref{lst:mp4box}.

\begin{lstlisting}[language=bash, caption={MP4Box command used to generate DASH manifest.}, label={lst:mp4box}]
MP4Box -dash 2000 -frag 2000 
       -profile dashavc264:live \
       -segment-name "segment_%s" 
       -out manifest.mpd \
       [lista_de_arquivos]
\end{lstlisting}

The main parameters defined in the command are described as follows:
\begin{itemize}
  \item \verb|-dash 2000|: Defines a segment duration of 2000~ms (2~seconds), establishing the temporal division of the video content.
  \item \verb|-frag 2000|: Enforces the same 2-second granularity for internal MP4 fragments, ensuring synchronization across all representations.
  \item \verb|-profile dashavc264:live|: Specifies the adopted DASH packaging profile, compatible with H.264/AVC encoding, typically used in live or on-demand streaming scenarios.
  \item \verb|-segment-name “segment_%s”|: Defines the naming convention for the generated segments (e.g., \texttt{segment\_1.m4s}, \texttt{segment\_2.m4s}, etc.). 
  \item \verb|-out manifest.mpd|: Sets the output filename for the DASH manifest, stored in XML format.
\end{itemize}

\subsection{Video Segment Acquisition}

The video segments delivered by the MPEG-DASH server are captured through \textit{HAR} (HTTP Archive) files generated directly by the browser during content playback. The \texttt{puppeteer-har} module executes this process by intercepting all HTTP requests from the browser and saving the collected data in a JSON-formatted file.

In contrast to tools such as \textit{tcpdump} and \textit{Wireshark}, which operate at the link or network layers, HAR-based capture functions at the application layer, enabling precise identification of individual DASH segment requests. This method records essential details, including URLs, file sizes, HTTP status codes, and response times, offering a fine-grained perspective on the OTT content delivery process.

The relevant video segments are extracted using the \texttt{jq} utility applied directly to the generated HAR files. The filtering criterion identifies URLs containing the string \texttt{/video-dash/segment\_bbb\_sunflower}, which corresponds to the MPEG-DASH video segments transmitted by the OTT server. The command below illustrates the filtering procedure:

\begin{lstlisting}[language=bash, caption={Filtering segments}, label={lst:har-jq}]
jq -r '.log.entries[] 
| select(.request.url 
| contains("http://IP-SERVER/video-dash/segment_
            bbb_sunflower")) 
| .request.url' <FILENAME>.har
\end{lstlisting}

From the extracted list of URLs, the video segments are automatically downloaded while preserving the original characteristics of each request, including the adaptive bitrate (quality) selected by the player during execution. This approach allows for the accurate reconstruction of video loading behavior under the different 5G FWA settings evaluated in the experiments. Preliminary tests with the \textit{Wireshark} and \textit{tcpdump} tools show inconsistent packet-level captures of video segments, preventing complete video reconstruction. In contrast, the HAR format, which operates at the application layer, reliably records all HTTP requests, including headers, latency, segment size, and temporal sequence, making it a more effective solution for this task and for potential QoE analysis of OTT content delivery.

\subsection{Video Reconstructions}

After the segment capture, the video reconstruction stage takes place. This process consists of three main steps:

\begin{enumerate}
    \item Group the segments into short fragments of the received video according to the resolution and bitrate of each segment.
    \item Harmonize the quality of the fragments by upscaling them to 4K. This step ensures that videos with lower quality than 4K visibly display their degradation.
    \item Concatenate all fragments with uniform quality to produce the degraded video.
\end{enumerate}

All of these steps use FFmpeg commands for execution.

\subsection{Video Metric Calculation}

FFmpeg supports the calculation of objective video quality metrics, including PSNR, SSIM, and VMAF. The filters \texttt{psnr}, \texttt{ssim}, and \texttt{libvmaf} allow for the comparison between the original (\texttt{Video-Original.mp4}) and the degraded (\texttt{Video-Degraded.mp4}) video files. The following commands describe the process for each metric:

\begin{itemize}
    \item \textbf{PSNR}: \texttt{ffmpeg -i Original-Video.mp4 -i Degraded-Video.mp4 -lavfi \\ "[0:v][1:v]psnr=stats\_file=psnr.txt\"" -f null} (calculate PSNR metric and writes in \texttt{psnr.txt} file). 
    \item \textbf{SSIM}: \texttt{ffmpeg -i Original-Video.mp4 -i Degraded-Video.mp4 -lavfi \\ "[0:v][1:v]ssim=stats\_file=ssim.txt\"" -f null} (calculate SSIM metric and writes in \texttt{ssim.txt} file). 
    \item \textbf{VMAF}: \texttt{ffmpeg -i Original-Video.mp4 -i Degraded-Video.mp4 -lavfi \\ "[0:v][1:v]libvmaf=log\_path=vmaf.xml\"" -f null} (calculate VMAF metric and writes in \texttt{vmaf.xml} file).

\end{itemize}

\subsection{Network Metrics Acquisition}

The \texttt{srsRAN} protocol stack~\cite{srsran2025} provides native support for reporting and visualizing Key Performance Indicator (KPI) metrics through Grafana, a monitoring platform that converts data from time-series databases (TSDBs) into interactive dashboards and visual representations. In the \texttt{srsRAN} codebase, this functionality is natively integrated through a Docker container located in the \texttt{docker} directory. The corresponding \texttt{docker-compose.yaml} file defines the services required for KPI collection and visualization. KPI data is exported to an InfluxDB database, which serves as the backend for Grafana dashboards.

The preconfigured dashboard allows per-UE monitoring of key network metrics, including Modulation and Coding Scheme (MCS), Signal-to-Noise Ratio (SNR) and Channel Quality Indicator (CQI), throughput, and the number of active UEs. This study focuses on observing the throughput behavior of the 5G FWA router, in order to quantify the relationship between network performance
and perceived visual quality of OTT content delivery. After each experimental campaign, the collected measurements are exported to \texttt{.csv} files for further analysis. Internal resources of the 5G FWA device, specifically its memory, CPU and Wi-Fi capacity, were monitored throughout the 120 experimental runs and remained consistently below saturation, ensuring that the 5G wireless link served as the primary bottleneck in the delivery chain.

\subsection{Proposed Campaigns Automation}
\label{sec:automation}

Given the complexity of the experimental setup, executing multiple campaigns with all parameter variations is both challenging and time-consuming. To overcome this limitation, we design an automated processing framework that integrates the entire workflow, from video preparation and 5G network configuration to final data collection. This section presents the proposed automation methodology.

The automation framework ensures synchronization, repeatability, and minimal manual intervention across all experimental runs. The approach controls the 5G link adaptation process, enabling the automatic selection of the highest modulation scheme, either up to 64-QAM or 256-QAM. In parallel, multiple client instances emulate DASH video consumption in a staggered and fully coordinated manner. All stages of execution are orchestrated via \texttt{cron}, to ensure precise timing and consistent experiment scheduling.

The \texttt{cron} utility, native to Unix-based systems, plays a central role in ensuring that events execute precisely according to schedule, following a predefined sequence. By orchestrating scheduled tasks across all machines, including client hosts and the 5G system server, the framework enables the execution of extended test sequences without requiring manual intervention or supervision.

%The experimental environment consisted of a base station with a USRP running \textit{gNB} via srsRAN and five Linux host machines connected via FWA access. Switching between the 64QAM and 256QAM modulation schemes was accomplished through \texttt{cron jobs} scheduled on the USRP server. Each 40-minute cycle, called a \textit{Job}, followed the sequence below:

As outlined earlier, the experimental setup consists of a 5G radio access network implemented with the srsRAN protocol stack on the Haykin node of the extreme edge domain of LANCE's telco-cloud testbed, which connects to a USRP RF front-end responsible for the over-the-air transmission of the 5G signal. The 5G FWA router distributes video data to five laptops via Wi-Fi. The transmission capacity of the 5G FWA link, configured for either 64-QAM or 256-QAM maximum modulation, is managed through \texttt{cron jobs} scheduled on the 5G server. Each 40-minute execution cycle, referred to as a \textit{Job}, follows the sequence described below:

\begin{enumerate}
    \item At time $t_0$, the gNB protocol stack (srsRAN) initializes with 64-QAM set as the maximum modulation for link adaptation.
    \item At time $t_0+5$, one client (laptop) automatically initiates DASH video playback.
    \item At time $t_0+10$, three clients simultaneously start DASH video playback.
    \item At time $t_0+15$, all five clients access DASH content concurrently.
    \item At time $t_0+20$, the gNB stack process is terminated, and the maximum modulation is switched to 256-QAM.
    \item From time $t_0+25$ to $t_0+40$, the same scaling pattern (1 $\rightarrow$ 3 $\rightarrow$ 5 users) is repeated, now with 256-QAM set as the maximum modulation for link adaptation.
\end{enumerate}

%On the clients, Node.js (\texttt{index.js})~\cite{ufrn2025dash} scripts based on the \texttt{puppeteer} and \texttt{puppeteer-har} (\texttt{package.json})~\cite{ufrn2025dash} libraries automated navigation and network data collection. Each run followed these steps:

On the client side, automation relied on Node.js (\texttt{index.js})~\cite{ufrn2025dash} script that used the \texttt{puppeteer} and \texttt{puppeteer-har} libraries, as specified in \texttt{package.json}~\cite{ufrn2025dash}, to control browser and capture network activity. Each execution followed the sequence below:

\begin{enumerate}
    \item Launch the Chrome browser in visible and \textbf{anonymous} mode, ensuring that no cache or stored data influences network performance measurements.
    \item Access the DASH server at \texttt{http://IP-SERVER}.
    \item Wait for the video player to load and start playback.
    \item Capture all network traffic in HAR format for a duration of 3 minutes.
    \item Save the HAR file with \textit{timestamp} in a dedicated directory for the corresponding job.
\end{enumerate}

All HAR files were organized by job and maximum modulation setting in directories such as \texttt{\textasciitilde/har\_logs/jobX/\allowbreak[64qam|256qam]} for subsequent centralized analysis.

Regarding the operational performance of the proposed framework, the automation was designed to ensure high reliability and scientific reproducibility with negligible system overhead. Because the Node.js and Puppeteer-based scripts operate exclusively at the application layer of the client stations, they do not interfere with the internal processing capacity or resource scheduling of the 5G core or RAN components. The framework’s reliability was validated through 120 independent execution without manual intervention, utilizing NTP synchronization to ensure temporal consistency. Furthermore, the architecture is orchestrated via deterministic system-level schedulers (cron jobs) across distributed stations, facilitating horizontal scalability. This design allows the experimental testbed to be expanded with additional concurrent probe nodes—efficiently deployable via configuration management tools such as Ansible—without requiring any modifications to the core automation logic. To ensure full reproducibility, the complete environment specification, including pinned dependency versions in the package.json and the orchestration scripts, is made publicly available in the paper’s repository~\cite{ufrn2025dash}.

\section{Result Discussion}
\label{sec:results}

Figures~\ref{fig:tput_5g},~\ref{fig:psnr},~\ref{fig:ssim}, and~\ref{fig:vmaf} present the average results from 10 independent experimental trials (jobs) conducted across the twelve examined scenarios. In all figures, subplots (a), (b), and (c) correspond to the SISO transmission mode, while subplots (d), (e), and (f) correspond to the 2$\times$2 MIMO configuration. The pairs (a,d), (b,e), and (c,f) represent results for one, three, and five connected users, respectively. The plot legends identify the MCS\textsubscript{max} configurations: one employing 64-QAM, with link adaptation supporting modulation orders of 2, 4, and 6 bits per symbol; and another using 256-QAM, which extends the modulation order to 8 bits per symbol in the highest MCS level.

%The results for 64-QAM and 256-QAM modulations of the highest MCS were plotted together in the graphs for each transmission mode configuration and number of connected users.

\subsection{Throughput Results and Discussions}
%Fig.~\ref{fig:tput_5g} displays the throughput variation (Mbps) on the vertical axis and the three-minute duration of the experiment (seconds) on the horizontal axis. Figs.~\ref{fig:tput_5g_1user_siso},~\ref{fig:tput_5g_3user_siso}, and~\ref{fig:tput_5g_5user_siso} show that when the gNB antennas are configured in SISO mode, modulation has minimal effect on system throughput. This contrasts with the results obtained for the MIMO configuration, where 256-QAM modulation provides higher throughput as the number of users connected to the FWA device increases. Considering the application's buffering capability, higher throughputs allow users to receive the whole video before the total duration of the experiment, as shown in Fig.~\ref{fig:tput_5g_5user_mimo}, where the curve for the 256-QAM configuration begins to decline after the 150th second, while the 64-QAM curve only starts to drop near the end of the experiment. 

\begin{figure*}[!ht]
    \centering
    \begin{subfigure}{0.33\linewidth}
        \centering
        \includegraphics[width=\linewidth]{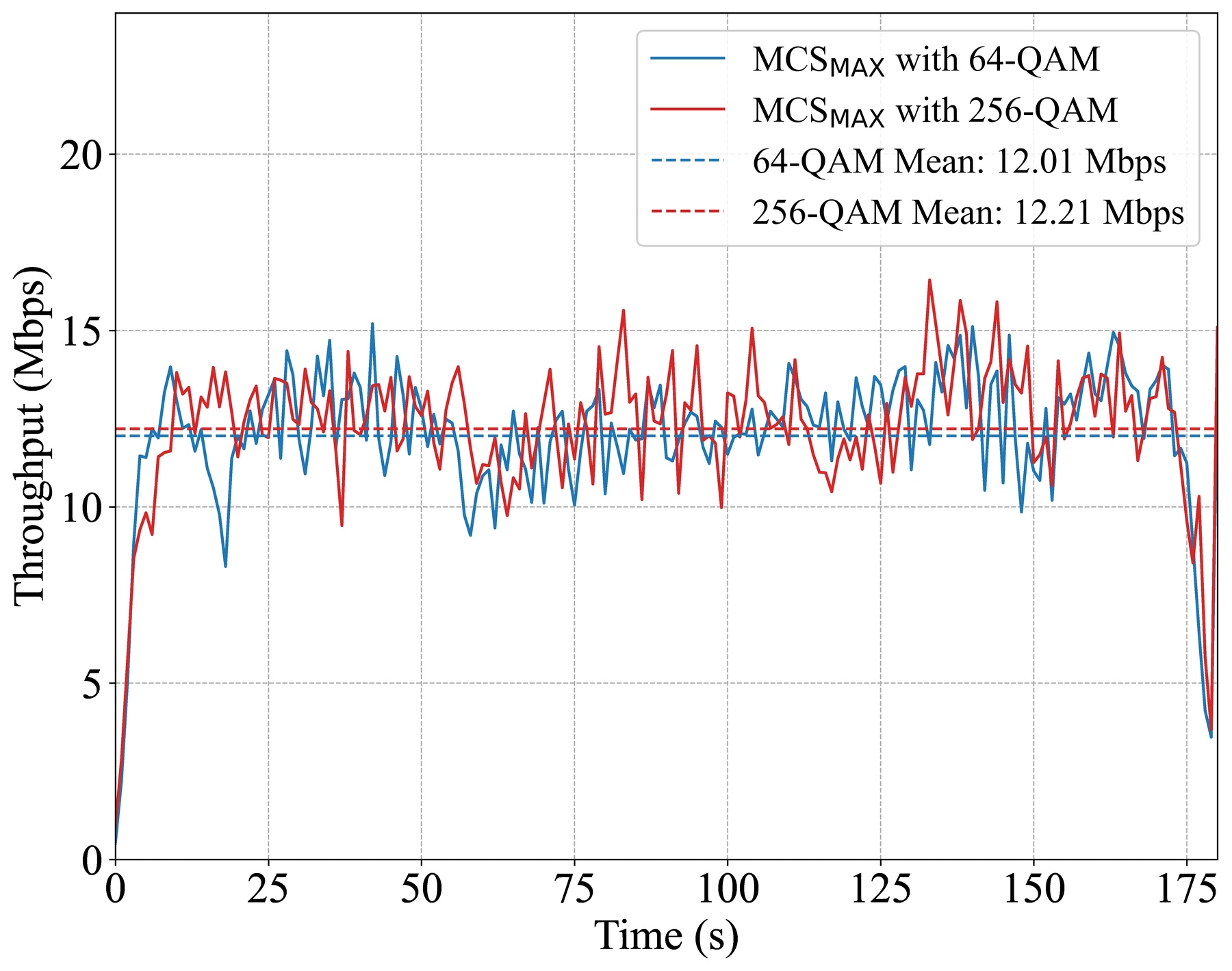}
        \caption{SISO - 1 User.}
        \label{fig:tput_5g_1user_siso}
    \end{subfigure}
    \begin{subfigure}{0.33\linewidth}
        \centering
        \includegraphics[width=\linewidth]{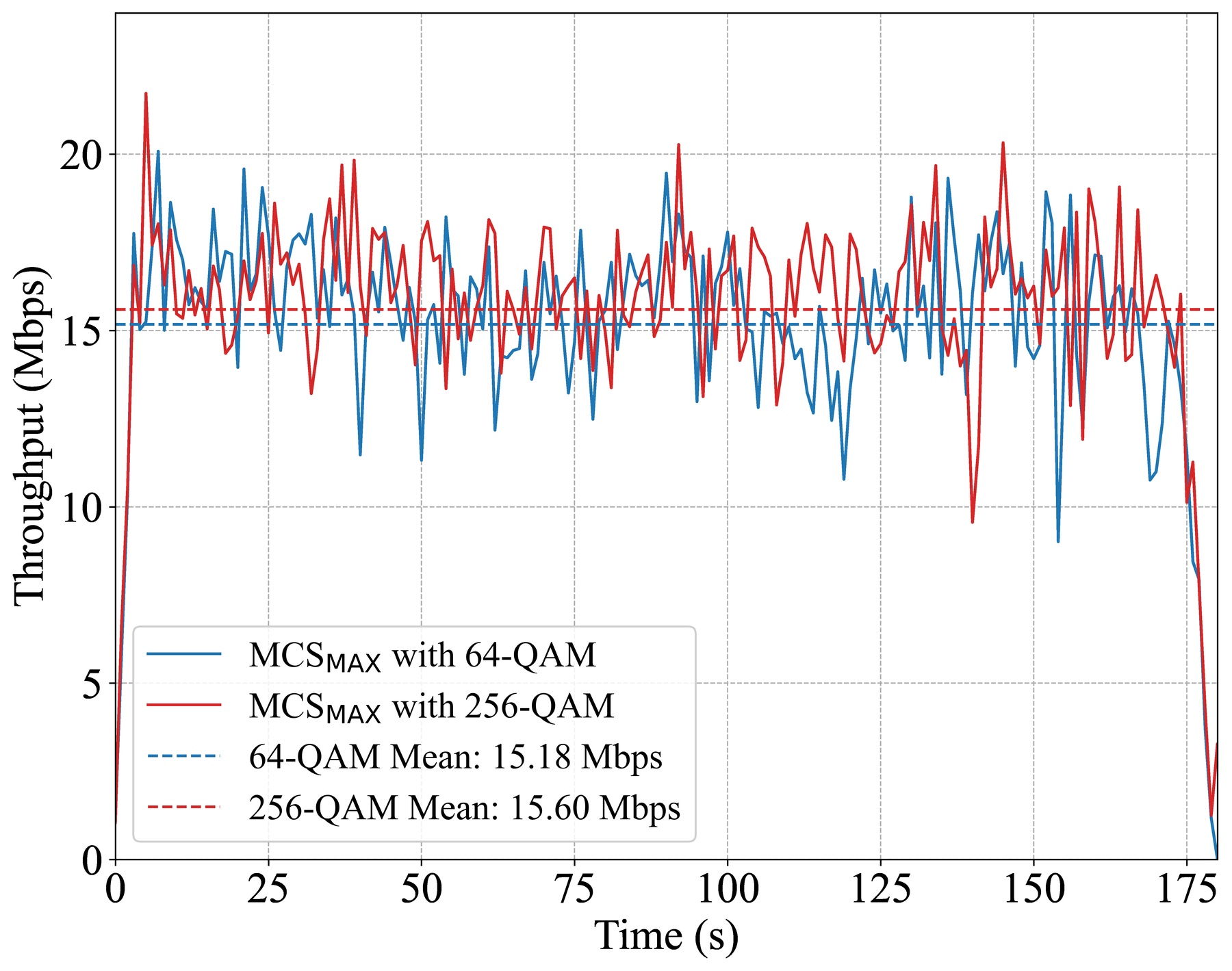}
        \caption{SISO - 3 Users.}
        \label{fig:tput_5g_3user_siso}
    \end{subfigure}
    \begin{subfigure}{0.33\linewidth}
        \centering
        \includegraphics[width=\linewidth]{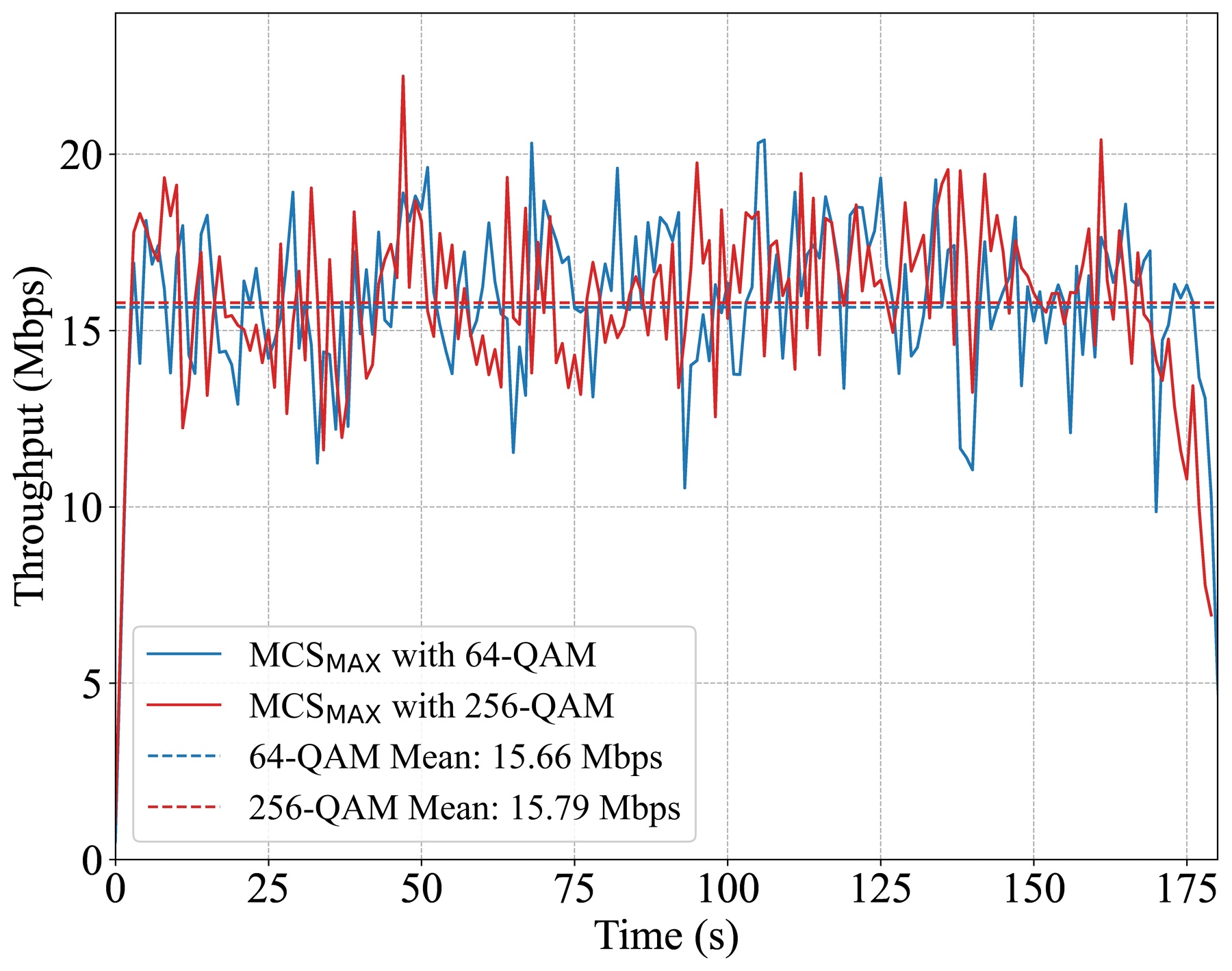}
        \caption{SISO - 5 Users.}
        \label{fig:tput_5g_5user_siso}
    \end{subfigure}
    \hfill
    \begin{subfigure}{0.33\linewidth}
        \centering
        \includegraphics[width=\linewidth]{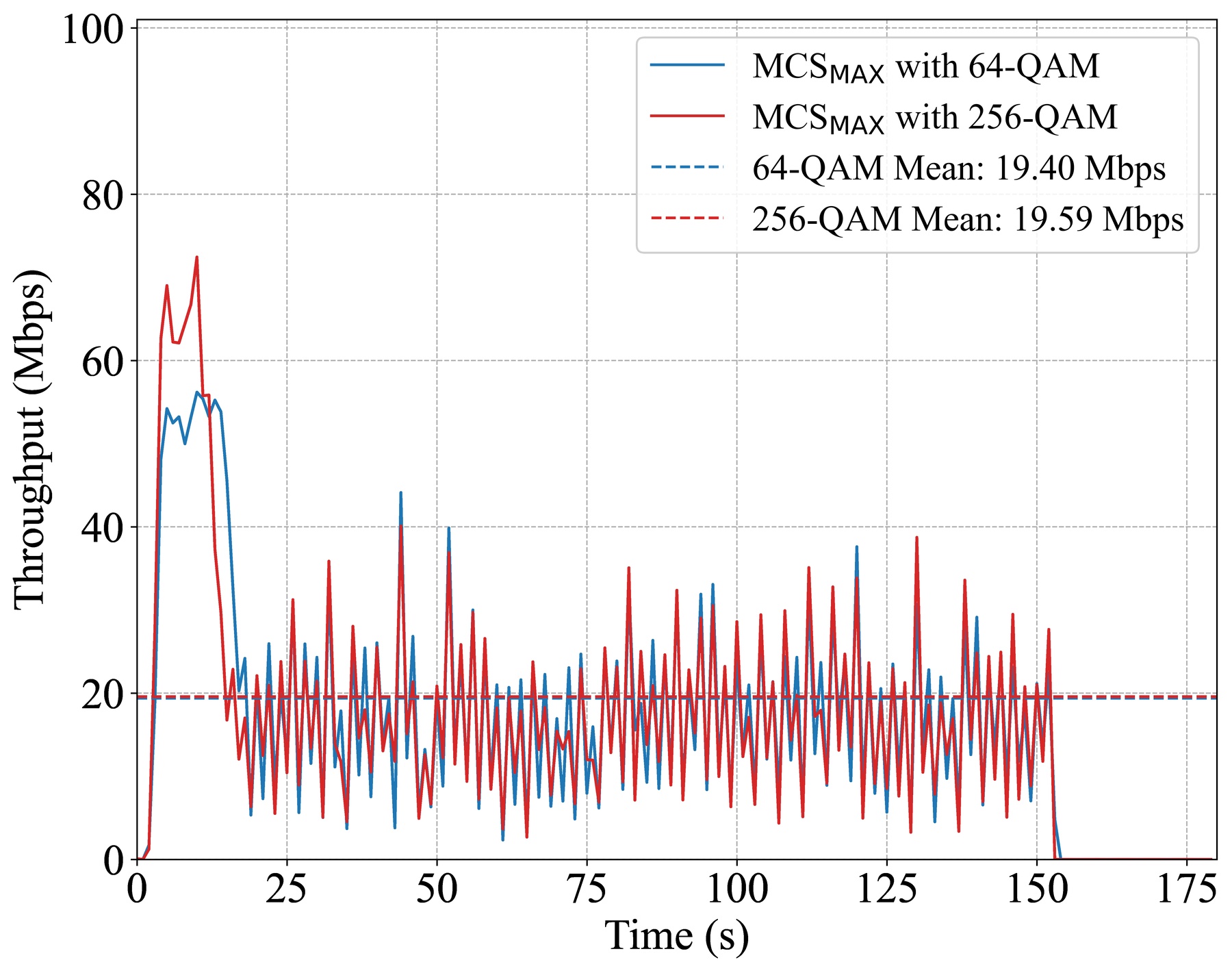}
        \caption{MIMO - 1 User.}
        \label{fig:tput_5g_1user_mimo}
    \end{subfigure}
    \begin{subfigure}{0.33\linewidth}
        \centering
        \includegraphics[width=\linewidth]{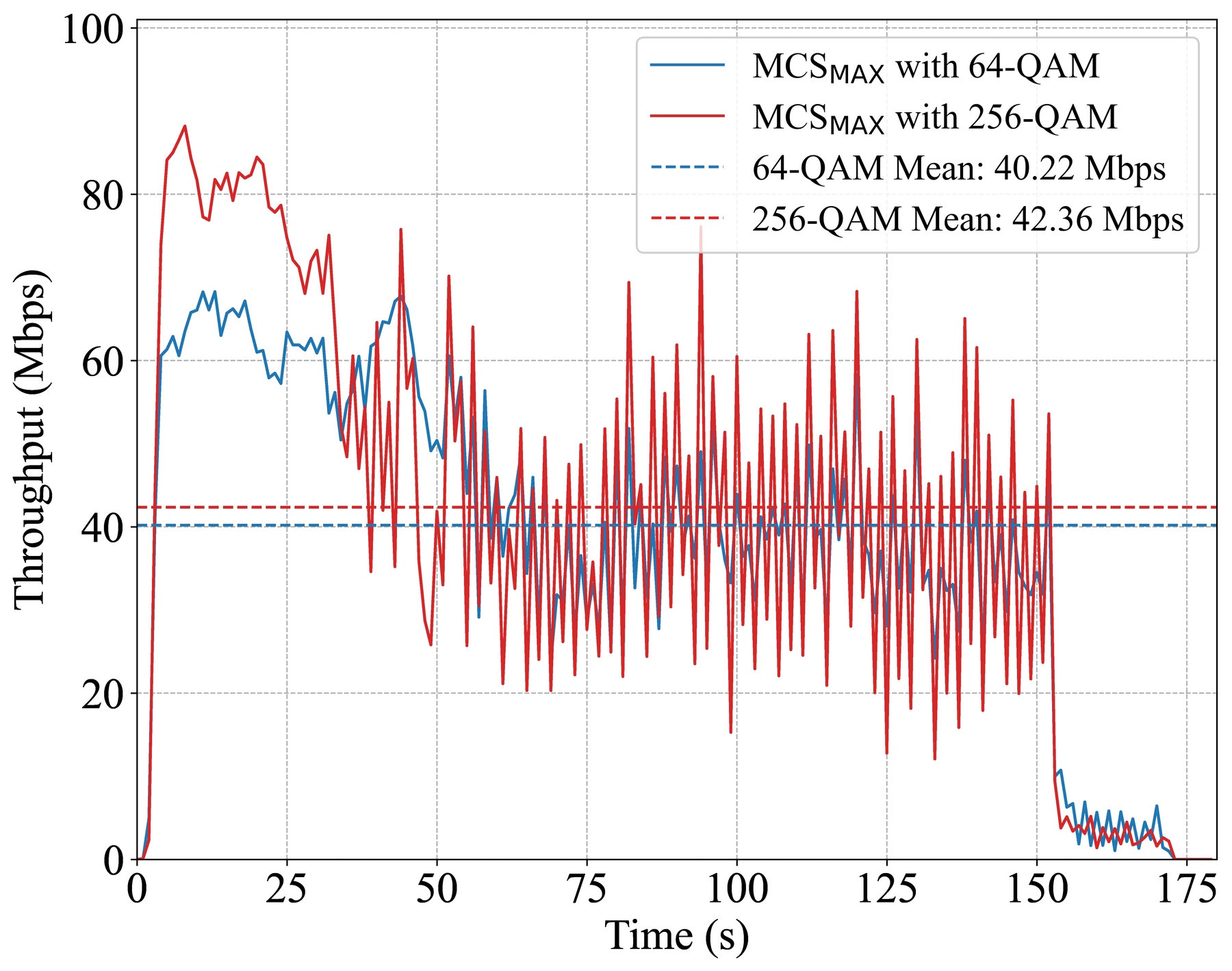}
        \caption{MIMO - 3 Users.}
        \label{fig:tput_5g_3user_mimo}
    \end{subfigure}
    \begin{subfigure}{0.33\linewidth}
        \centering
        \includegraphics[width=\linewidth]{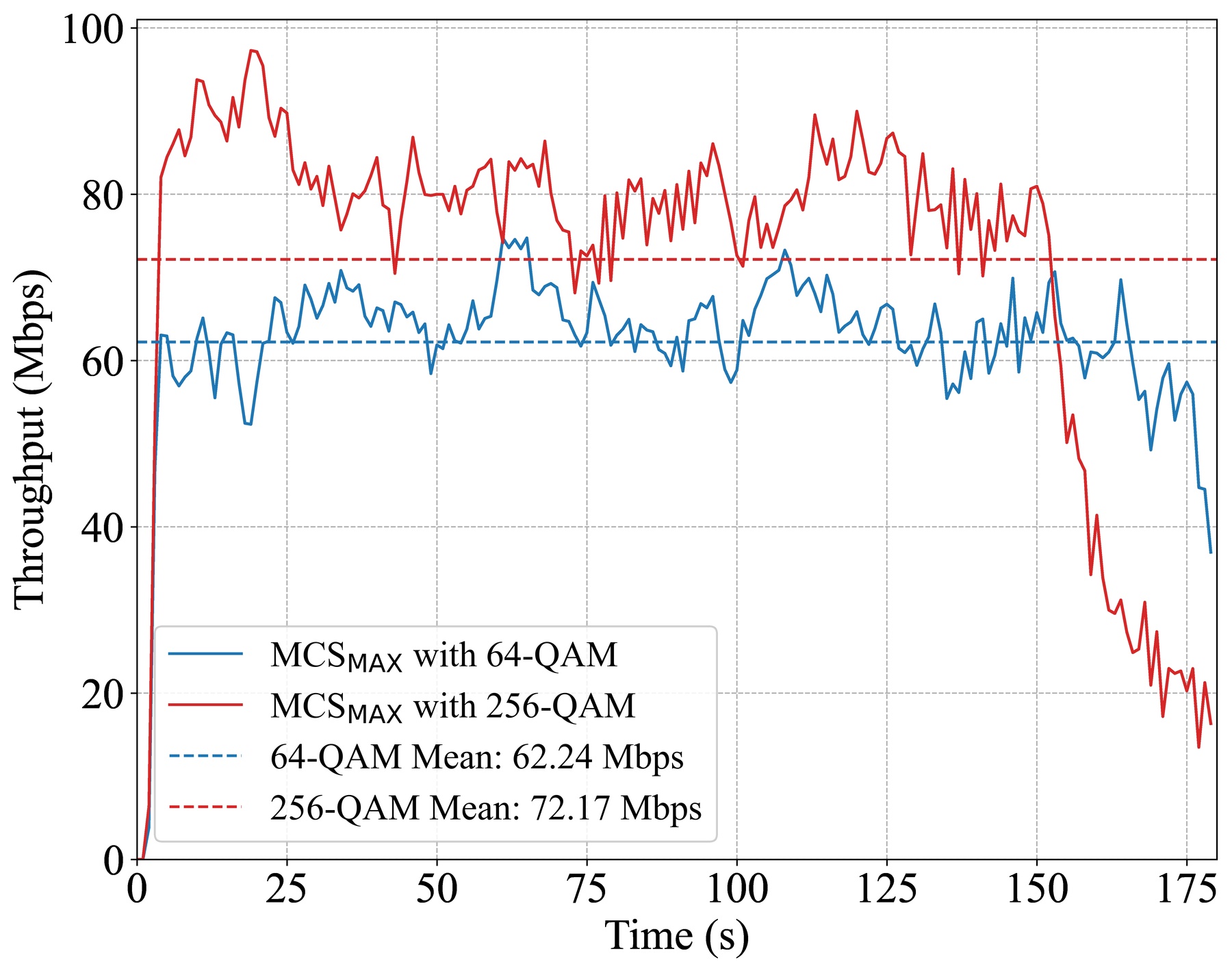}
        \caption{MIMO - 5 Users.}
        \label{fig:tput_5g_5user_mimo}
    \end{subfigure}
    \caption{Time evolution of the FWA link throughput (Mbps) during the experiment.}
    \label{fig:tput_5g}
\end{figure*}

Figure~\ref{fig:tput_5g} shows the measured throughput (in Mbps) as a function of time (in seconds) throughout the three-minute experiment. The initial peak corresponds to the video fetching phase, where segments are buffered to ensure smooth playback. Subfigures~\ref{fig:tput_5g_1user_siso},~\ref{fig:tput_5g_3user_siso}, and~\ref{fig:tput_5g_5user_siso} indicate that, under the SISO transmission mode, throughput remains relatively stable and shows minimal sensitivity to variations in the maximum modulation (MCS\textsubscript{max}). In contrast, the MIMO configuration demonstrates a clear performance distinction: the 256-QAM MCS\textsubscript{max} consistently achieves higher throughput as the number of users connected to the FWA router increases. Owing to the OTT application's buffering capability, higher throughput enables users to download the entire video before the experiment concludes, as observed in Subfigure~\ref{fig:tput_5g_5user_mimo}. The 256-QAM MCS\textsubscript{max} curve exhibits a gradual decline after approximately 150 seconds, whereas the 64-QAM MCS\textsubscript{max} curve begins to drop only near the end of the three-minute window.

%While in the SISO configuration the average throughput plateaued at 15 Mbps when more than three users were connected, in the MIMO configuration, the rates increased with the number of users. This may be due to the nature of video streaming traffic, which utilizes only the necessary bandwidth for the application. The SISO configuration's reduced bandwidth compared to the MIMO suggests that the SISO connection may have been a bottleneck with five users, whereas the MIMO can still handle more users.

In the SISO configuration, the average throughput stabilized around 15 Mbps once more than three users were connected. In contrast, the MIMO configuration exhibited a clear increase in throughput as the number of users grew. This difference stems from the adaptive nature of the OTT streaming traffic, which utilizes only the bandwidth necessary for continuous playback. The limited capacity in the SISO setup likely introduced a bottleneck under heavier load, particularly with five concurrent users, whereas the MIMO configuration provided sufficient spatial multiplexing to accommodate additional users without noticeable degradation in throughput.

\subsection{Video Metrics Results and Discussions}

%Figure~\ref{fig:psnr} shows the PSNR results, with the vertical axis displaying the PSNR (in dB) and the horizontal axis indicating the video frames, with a total of 10,800 frames received over three minutes.

\begin{figure*}[!ht]
    \begin{subfigure}{0.33\linewidth}
        \centering
        \includegraphics[width=\linewidth]{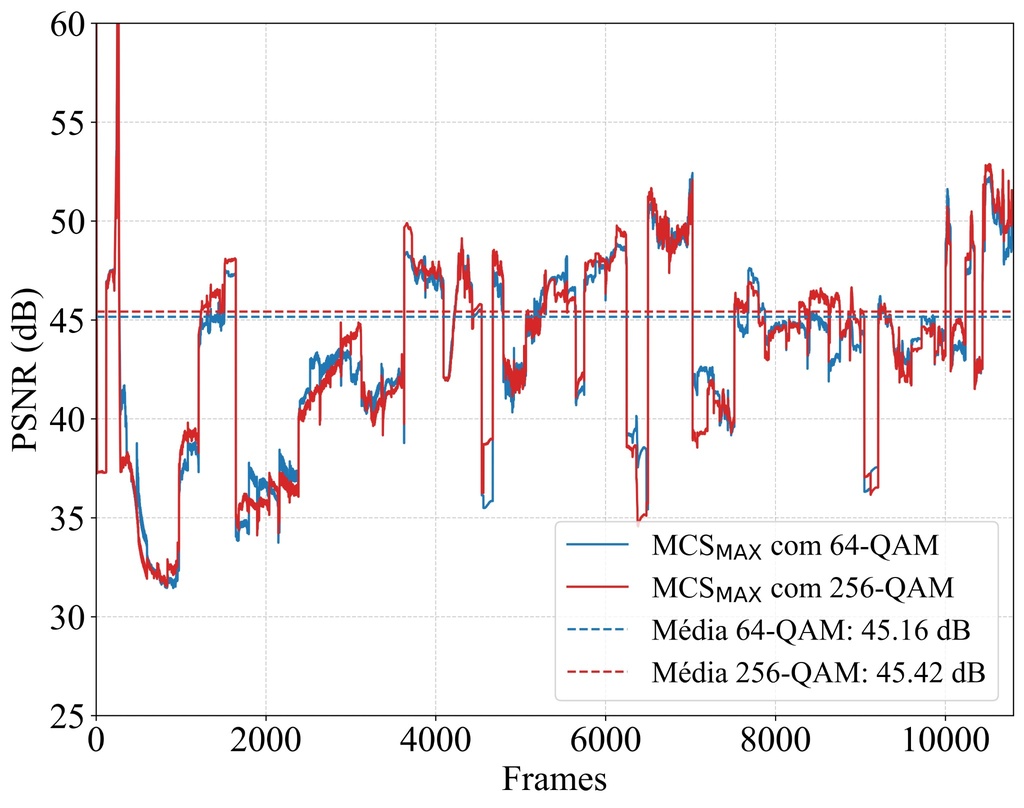}
        \caption{SISO - 1 User.}
        \label{fig:psnr_1user_siso}
    \end{subfigure}
    \begin{subfigure}{0.33\linewidth}
        \centering
        \includegraphics[width=\linewidth]{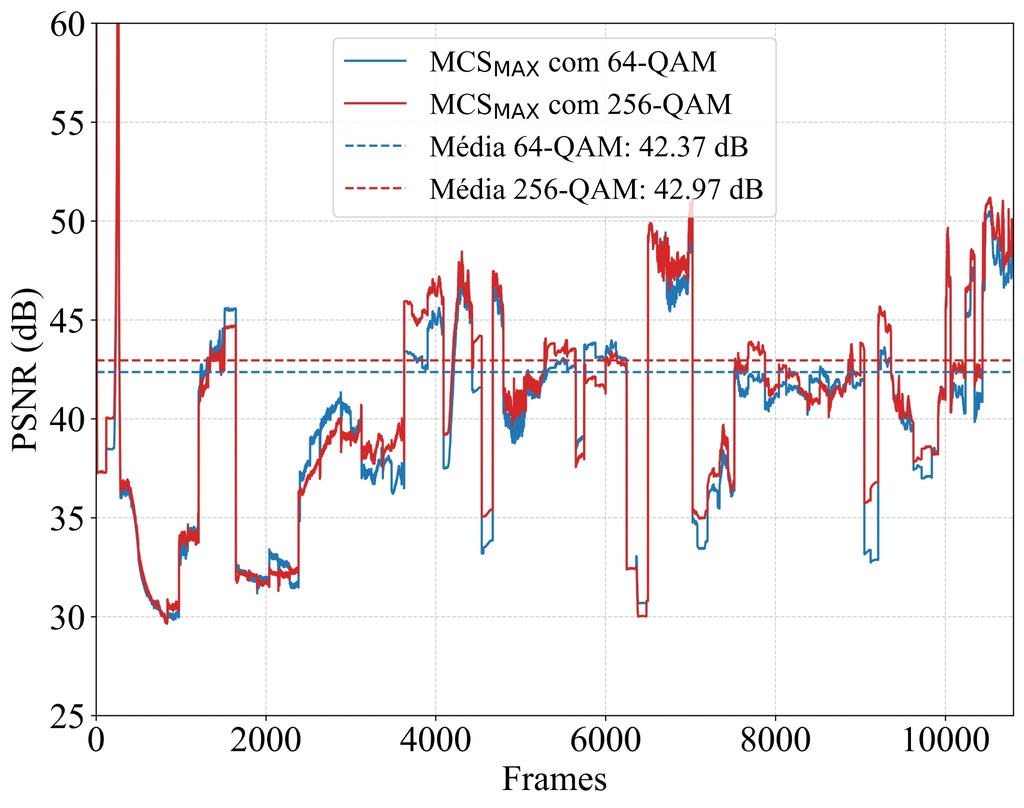}
        \caption{SISO - 3 Users.}
        \label{fig:psnr_3user_siso}
    \end{subfigure}
    \begin{subfigure}{0.33\linewidth}
        \centering
        \includegraphics[width=\linewidth]{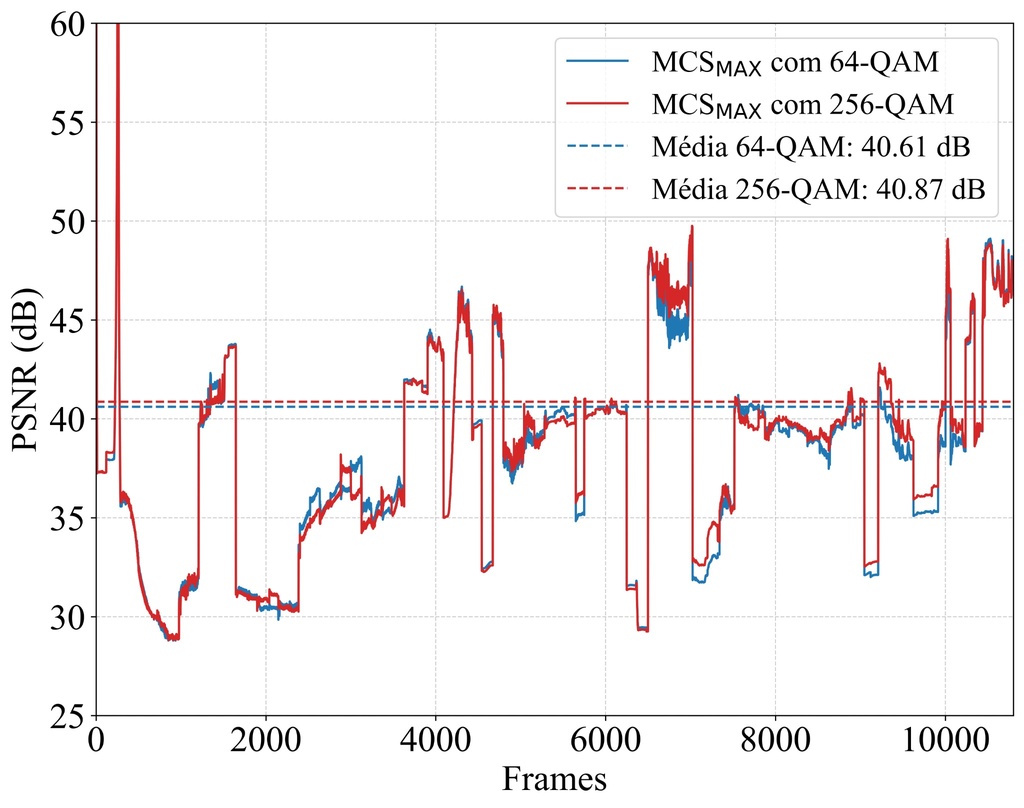}
        \caption{SISO - 5 Users.}
        \label{fig:psnr_5user_siso}
    \end{subfigure}
    \hfill
    \begin{subfigure}{0.33\linewidth}
        \centering
        \includegraphics[width=\linewidth]{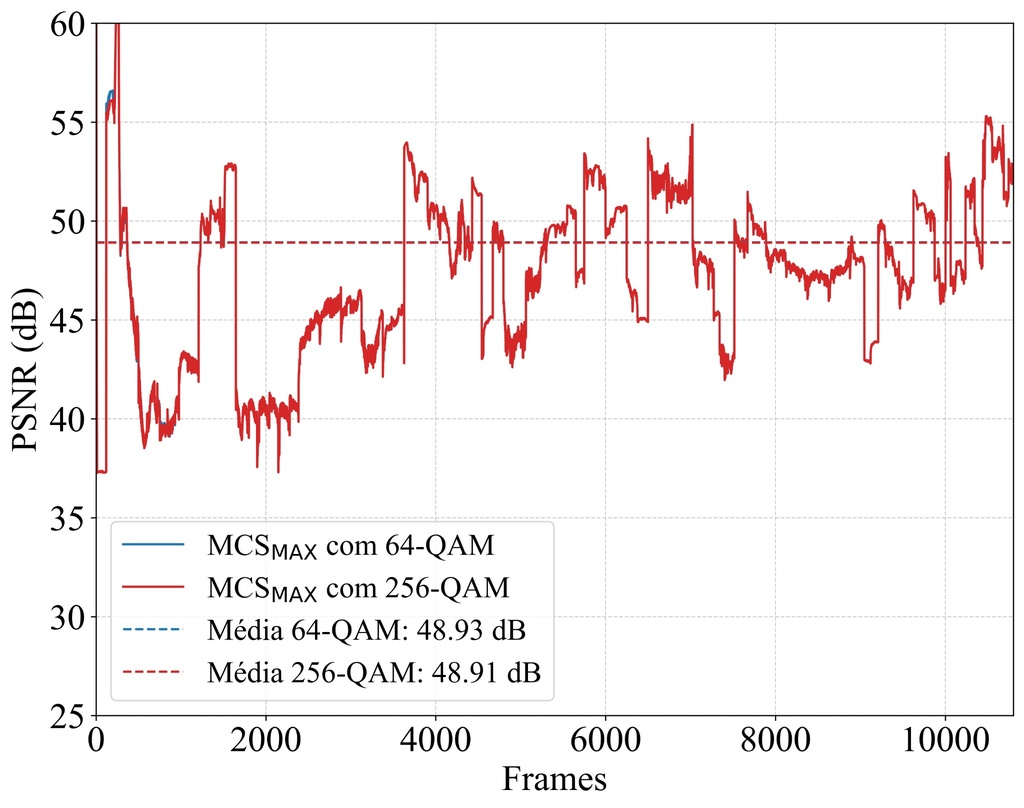}
        \caption{MIMO - 1 User.}
        \label{fig:psnr_1user_mimo}
    \end{subfigure}
    \begin{subfigure}{0.33\linewidth}
        \centering
        \includegraphics[width=\linewidth]{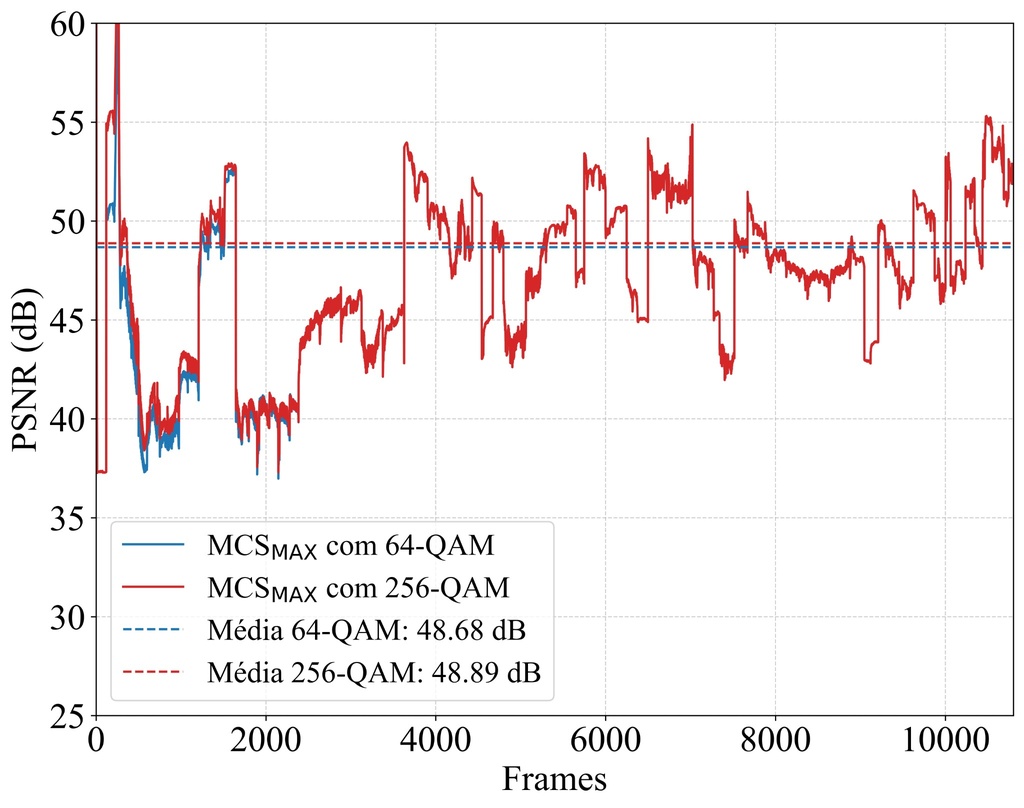}
        \caption{MIMO - 3 Users.}
        \label{fig:psnr_3user_mimo}
    \end{subfigure}
    \begin{subfigure}{0.33\linewidth}
        \centering
        \includegraphics[width=\linewidth]{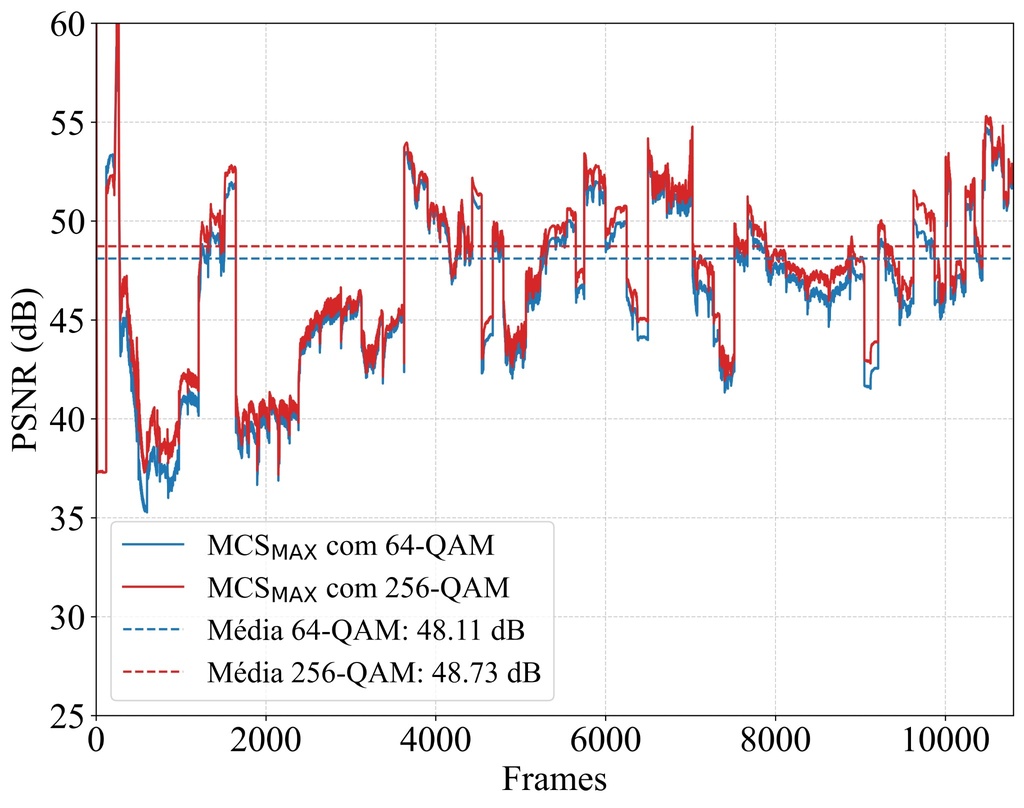}
        \caption{MIMO - 5 Users.}
        \label{fig:psnr_5user_mimo}
    \end{subfigure}
    \caption{Variation of the PSNR (dB) across the received frames.}
    \label{fig:psnr}
\end{figure*}

Figure~\ref{fig:psnr} presents the PSNR values, in decibels, as a function of the 10,800 video frames captured during the three-minute experiment. A clear trend emerges in which higher network loads and lower maximum modulation orders result in reduced PSNR values. Nonetheless, all measurements remain above 35~dB, corresponding to good perceptual video quality.

Table~\ref{tab:diferencas} summarizes the average PSNR values across the full experiment duration and reports the differences between the MIMO and SISO configurations. Consistent with the throughput results, the SISO configuration yields lower mean PSNR values than MIMO. The observed gain with MIMO ranges from 3.49 to 7.86~dB, highlighting the significant improvement in perceptual video quality enabled by the multi-antenna configuration.

\begin{table}[!ht]
\small
\caption{Difference between MIMO and SISO PSNR results per number of users.}
\label{tab:diferencas}
\centering
\begin{tabular}{c|c|c}
 \hline
 \multicolumn{1}{c|}{\textbf{Number of Users}} & \multicolumn{1}{c}{\textbf{64-QAM}} & \multicolumn{1}{c}{\textbf{256-QAM}} \\
 \hline
 1 & 3.77 dB & 3.49 dB\\
 3 & 6.31 dB & 5.92 dB\\
 5 & 7.50 dB & 7.86 dB\\
 \hline
\end{tabular}
\end{table}

% PSNR degradation intensifies more rapidly with increasing network load in the SISO configuration compared to the MIMO configuration.

PSNR degradation progresses more rapidly with increasing network load in the SISO configuration than in the MIMO configuration. The gap between the MIMO and SISO mean PSNR values widens as the number of connected users increases. When one or three users are active, the difference between the mean values is slightly smaller for MCS\textsubscript{max} with 256-QAM. However, with five concurrent users, the MIMO configuration under MCS\textsubscript{max} with 256-QAM yields a larger improvement compared to 64-QAM. Overall, the maximum MCS modulation exerts only a marginal influence on PSNR, as the mean values for 64-QAM and 256-QAM remain closely aligned across all scenarios, with a slight and consistent advantage for MCS\textsubscript{max} with 256-QAM, as shown in Figure~\ref{fig:psnr}.

%Figs.~\ref{fig:psnr_1user_siso}, ~\ref{fig:psnr_5user_siso} show that in the SISO configuration, the PSNR difference between the 1-user and 5-user scenarios is 4.65~dB for 64-QAM and 4.57~dB for 256-QAM. In the MIMO configuration, Figs.~\ref{fig:psnr_1user_mimo} and~\ref{fig:psnr_5user_mimo} show a difference of 1.64~dB for 64-QAM and just 0.34~dB for 256-QAM. The maximum MCS modulation does not significantly impact the SISO configuration, whereas the 256-QAM modulation maintains video quality more consistently in the MIMO configuration.

%Fig.~\ref{fig:ssim} shows the results for SSIM metric in the vertical axis, varying between 0 and 1, and the horizontal axis indicates each frame received over three minutes. 

\begin{figure*}[!ht]
    \begin{subfigure}{0.33\linewidth}
         \centering
         \includegraphics[width=\linewidth]{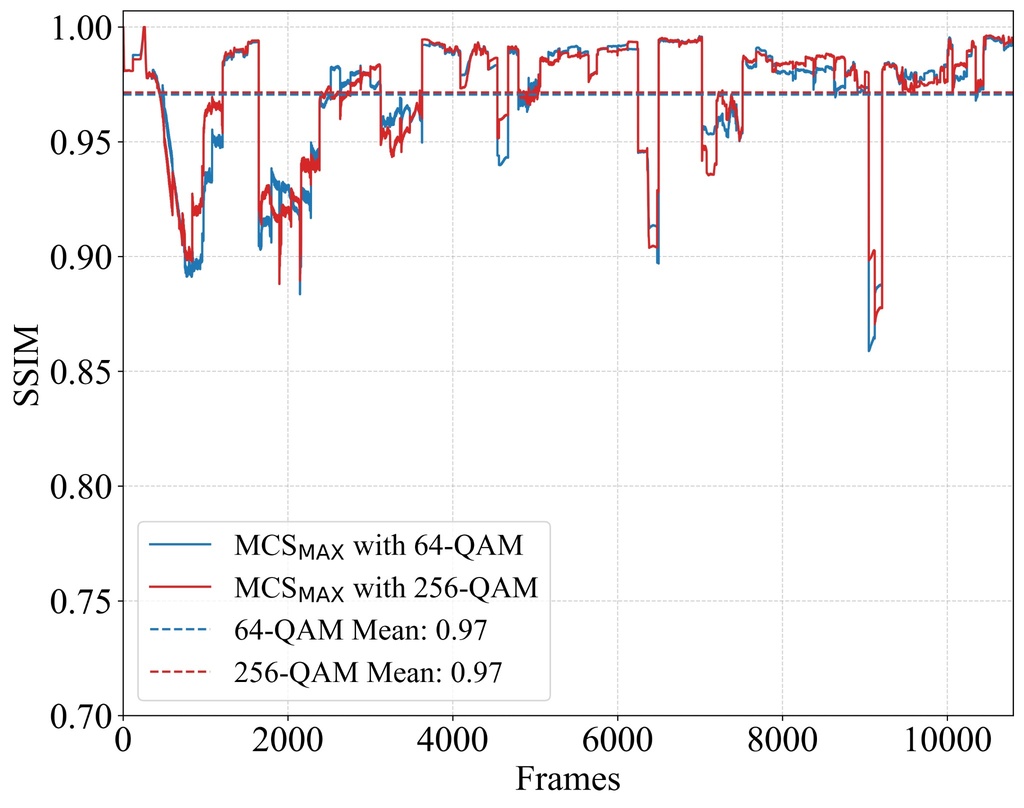}
         \caption{SISO - 1 User.}
         \label{fig:ssim_1user_siso}
    \end{subfigure}
     \begin{subfigure}{0.33\linewidth}
        \centering
        \includegraphics[width=\linewidth]{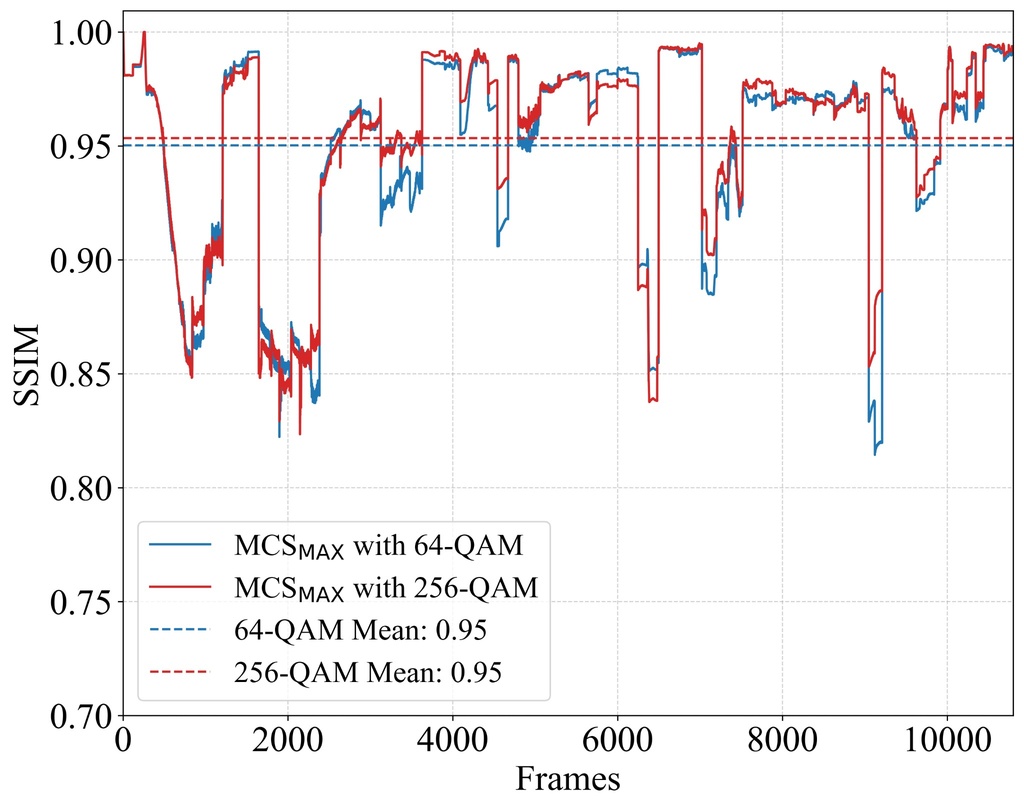}
        \caption{SISO - 3 Users.}
        \label{fig:ssim_3user_siso}
    \end{subfigure}
    \begin{subfigure}{0.33\linewidth}
        \centering
        \includegraphics[width=\linewidth]{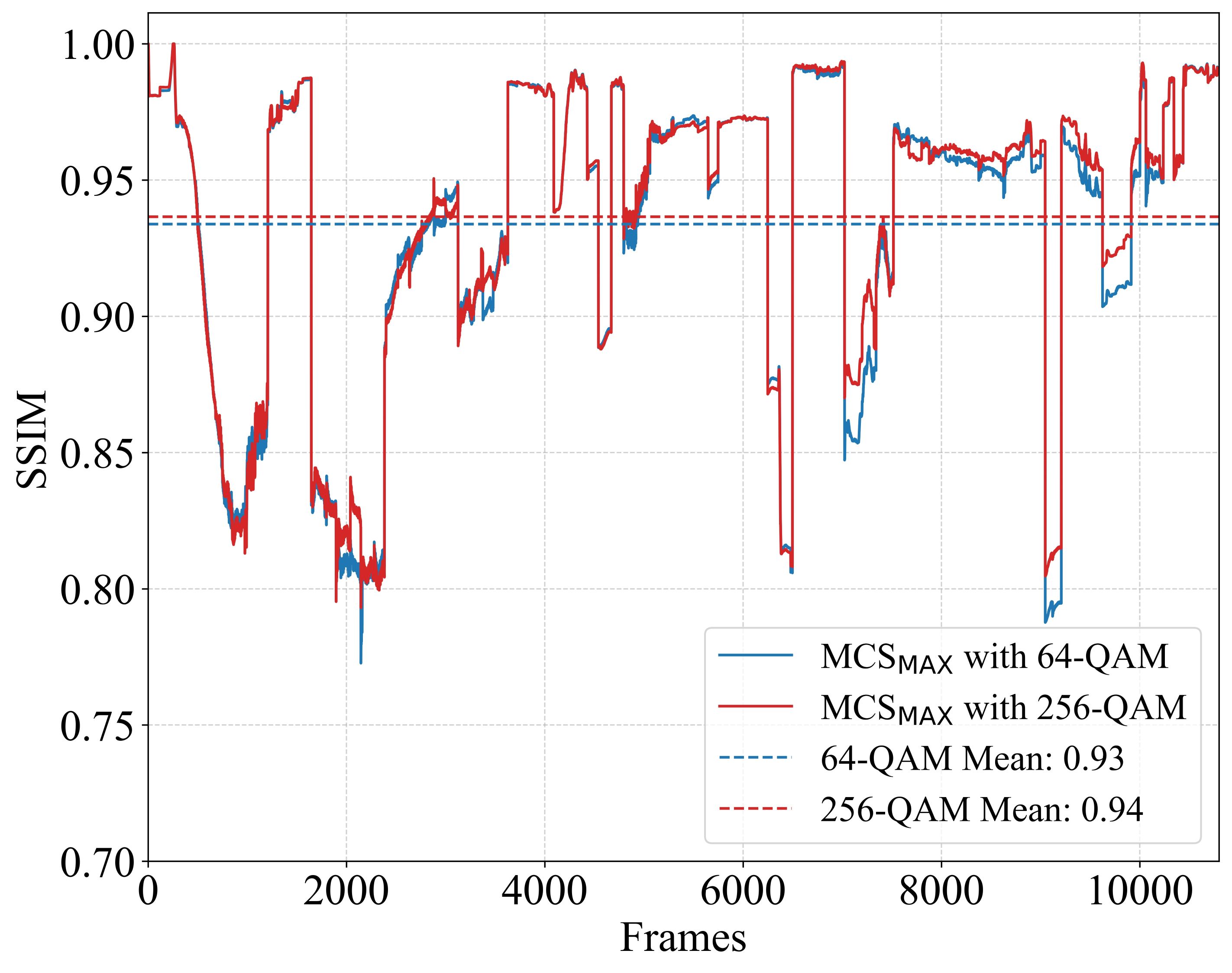}
        \caption{SISO - 5 Users.}
        \label{fig:ssim_5user_siso}
    \end{subfigure}
    \hfill
    \begin{subfigure}{0.33\linewidth}
        \centering
        \includegraphics[width=\linewidth]{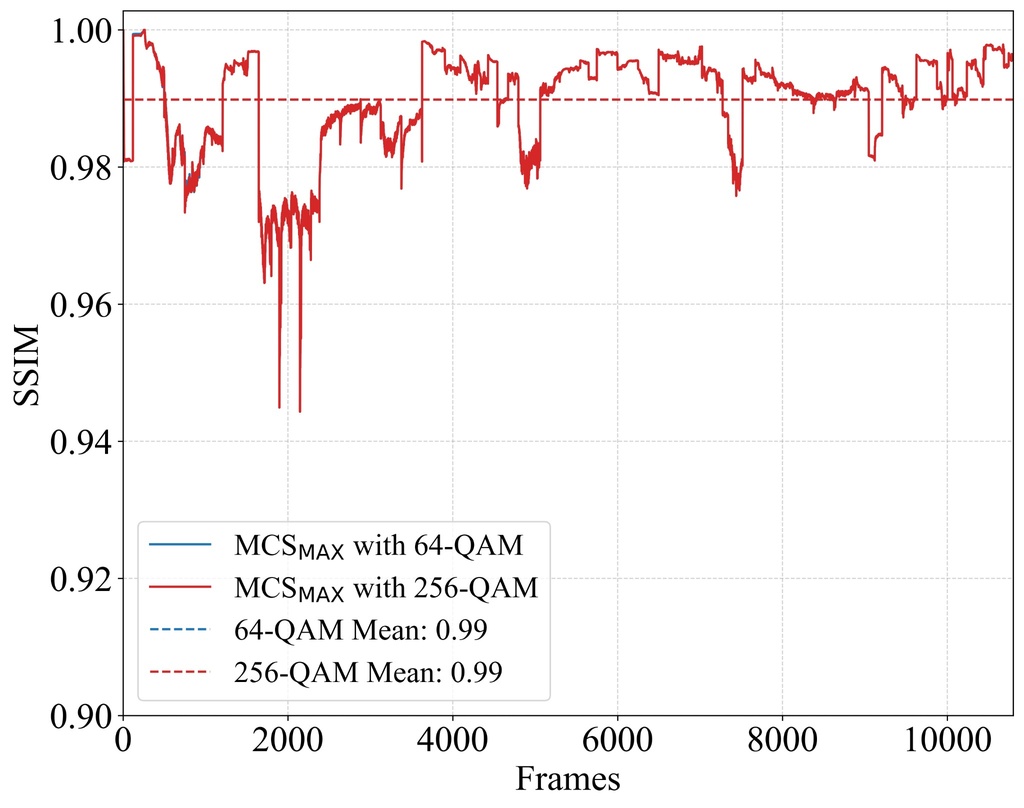}
        \caption{MIMO - 1 User.}
        \label{fig:ssim_1user_mimo}
    \end{subfigure}
    \begin{subfigure}{0.33\linewidth}
        \centering
        \includegraphics[width=\linewidth]{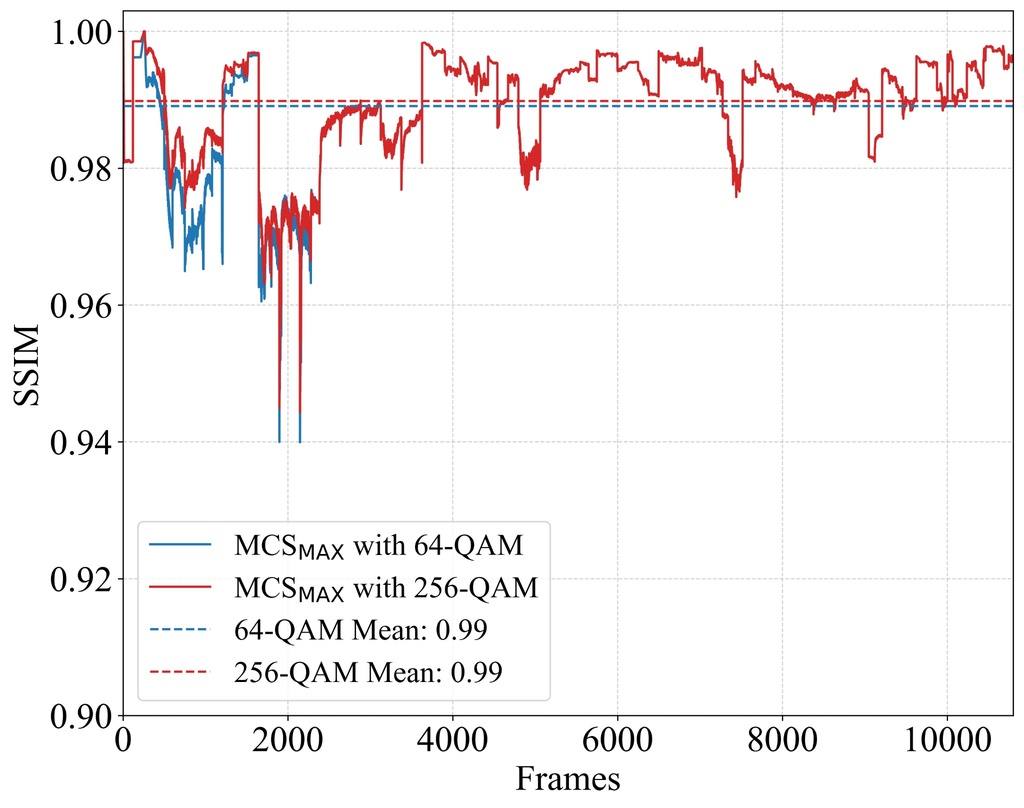}
        \caption{MIMO - 3 Users.}
        \label{fig:ssim_3user_mimo}
    \end{subfigure}
    \begin{subfigure}{0.33\linewidth}
        \centering
        \includegraphics[width=\linewidth]{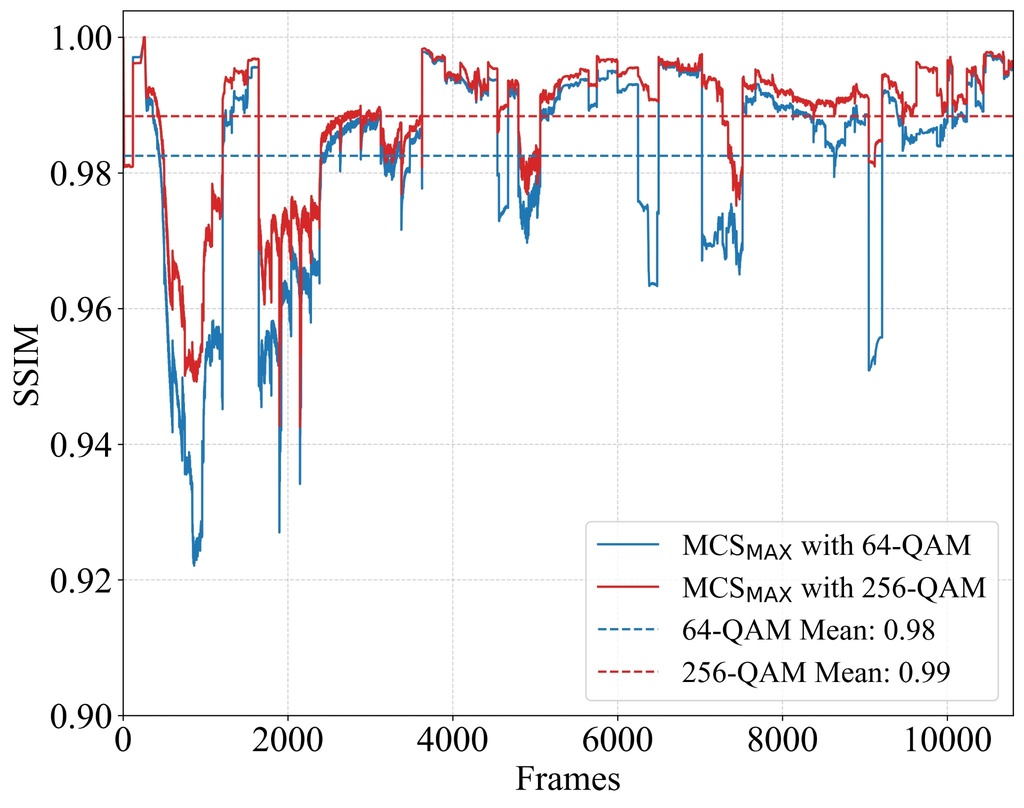}
        \caption{MIMO - 5 Users.}
        \label{fig:ssim_5user_mimo}
    \end{subfigure}
    \caption{Variation of the SSIM across the received frames.}
    \label{fig:ssim}
\end{figure*}

Figure~\ref{fig:ssim} presents the SSIM values (ranging from 0 to 1) computed over the OTT video frames received during the three-minute experiment. For the SISO transmission mode (Subfigures~\ref{fig:ssim_1user_siso},~\ref{fig:ssim_3user_siso} and~\ref{fig:ssim_5user_siso}), the mean SSIM decreases as the number of connected users increases, following the same trend observed in the PSNR results. In contrast, the MIMO configuration (Subfigures~\ref{fig:ssim_1user_mimo},~\ref{fig:ssim_3user_mimo} and~\ref{fig:ssim_5user_mimo}) maintains an average SSIM of around 0.99, indicating that the reconstructed frames are nearly identical to the originals. Overall, all MIMO scenarios deliver visual quality that is indistinguishable or perceptually identical to the reference video, while the SISO configurations produce occasional frames with minor visible differences, though still within an acceptable range for most users.

%The results using the SISO transmission mode (Figures~\ref{fig:ssim_1user_siso},~\ref{fig:ssim_3user_siso} and~\ref{fig:ssim_5user_siso}) demonstrated a drop in the SSIM mean as the number of connected users increased, which is consistent with the pattern shown in the PSNR results. With SSIM means of roughly 0.99, the MIMO configuration (Figures~\ref{fig:ssim_1user_mimo},~\ref{fig:ssim_3user_mimo} and~\ref{fig:ssim_5user_mimo}) produced frames that were highly compatible with the originals. %In all experiments, a drop in frame similarity occurred around the 2000th frame. %momento em que o DASH diminuiu a qualidade de vídeo. 
%Around the 9,000th frame, another decline was also observed in the SISO-configured scenarios. %Isso pode ter se dado pela...

%Finally, Figure~\ref{fig:vmaf} shows the results for VMAF score (in percentage) in the vertical axis and each frame transmitted in the experiment in the horizontal axis.
%Finally, 

\begin{figure*}[!ht]
    \centering
    \begin{subfigure}{0.33\linewidth}
        \centering
        \includegraphics[width=\linewidth]{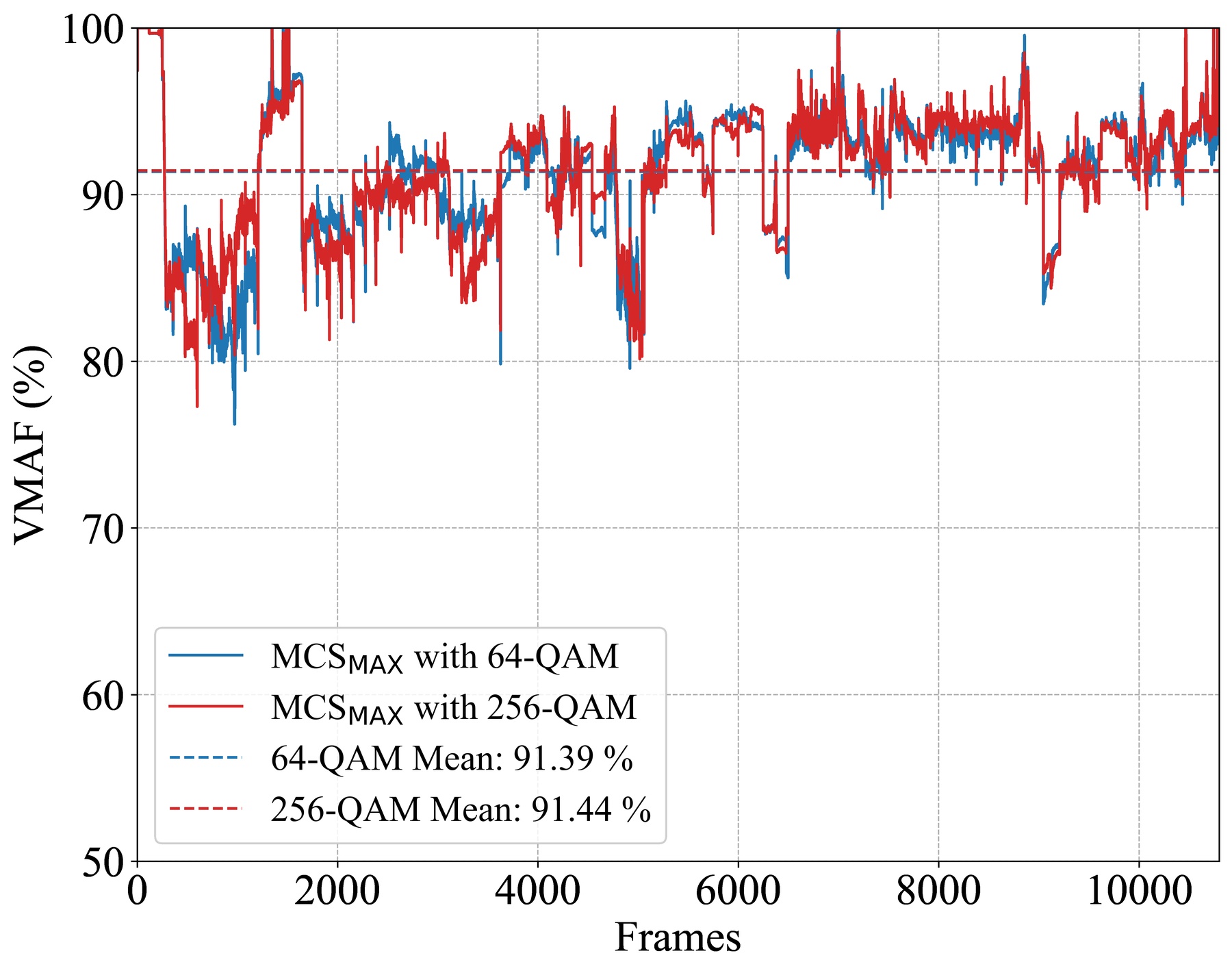}
        \caption{SISO - 1 User.}
        \label{fig:vmaf_1user_siso}
    \end{subfigure}
    \begin{subfigure}{0.33\linewidth}
        \centering
        \includegraphics[width=\linewidth]{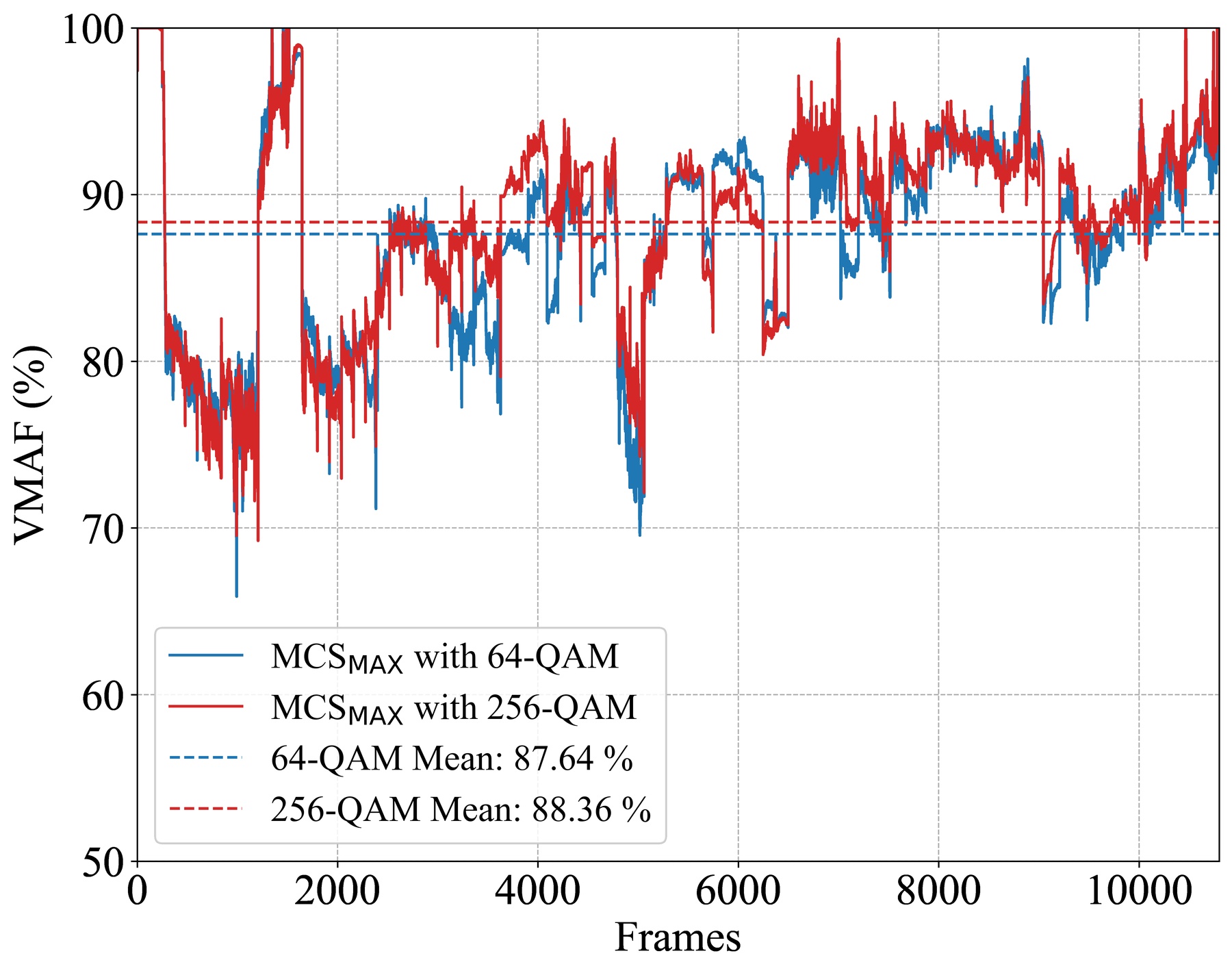}
        \caption{SISO - 3 Users.}
        \label{fig:vmaf_3user_siso}
    \end{subfigure}
    \begin{subfigure}{0.33\linewidth}
        \centering
        \includegraphics[width=\linewidth]{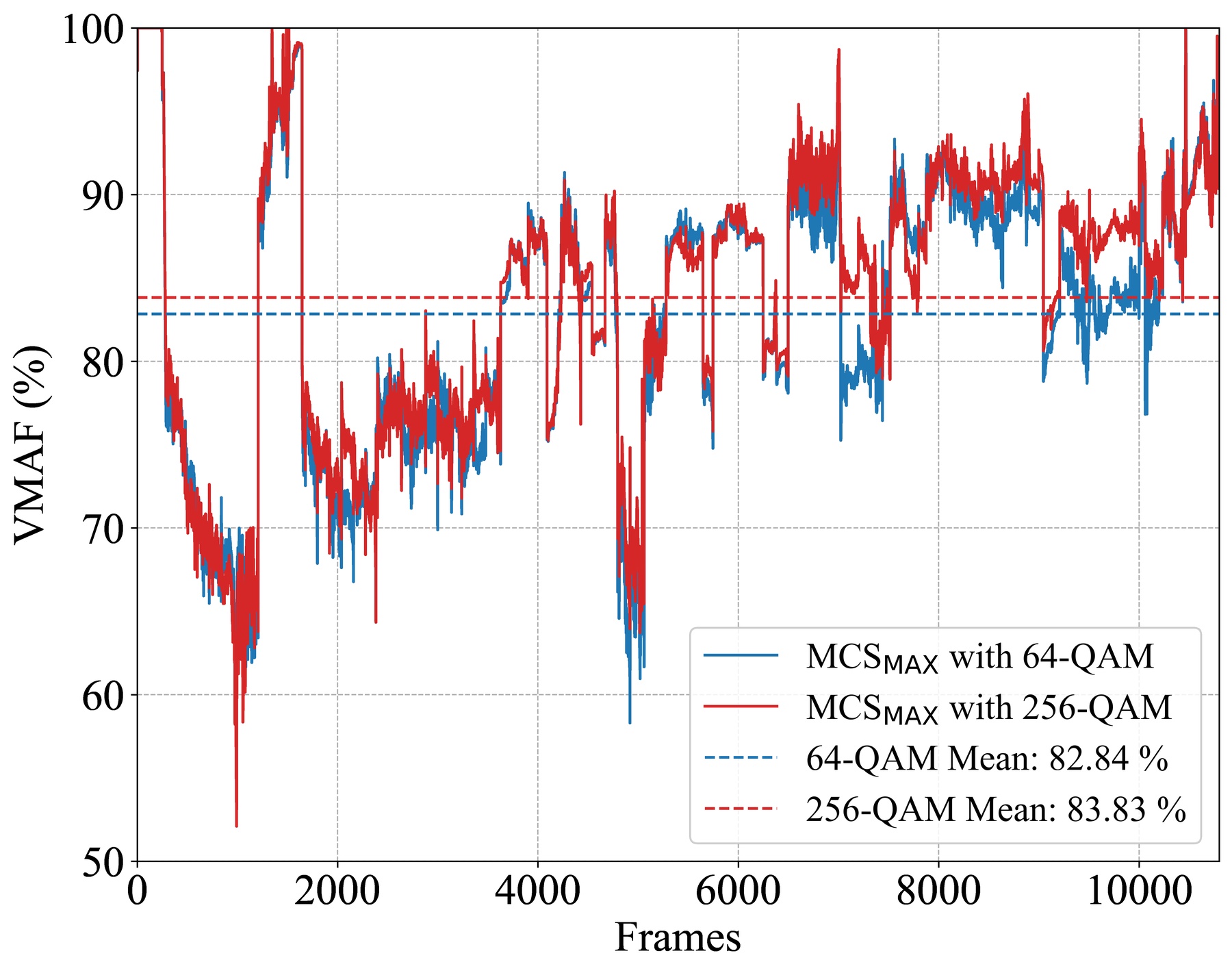}
        \caption{SISO - 5 Users.}
       \label{fig:vmaf_5user_siso}
    \end{subfigure}
    \hfill
    \begin{subfigure}{0.33\linewidth}
        \centering
        \includegraphics[width=\linewidth]{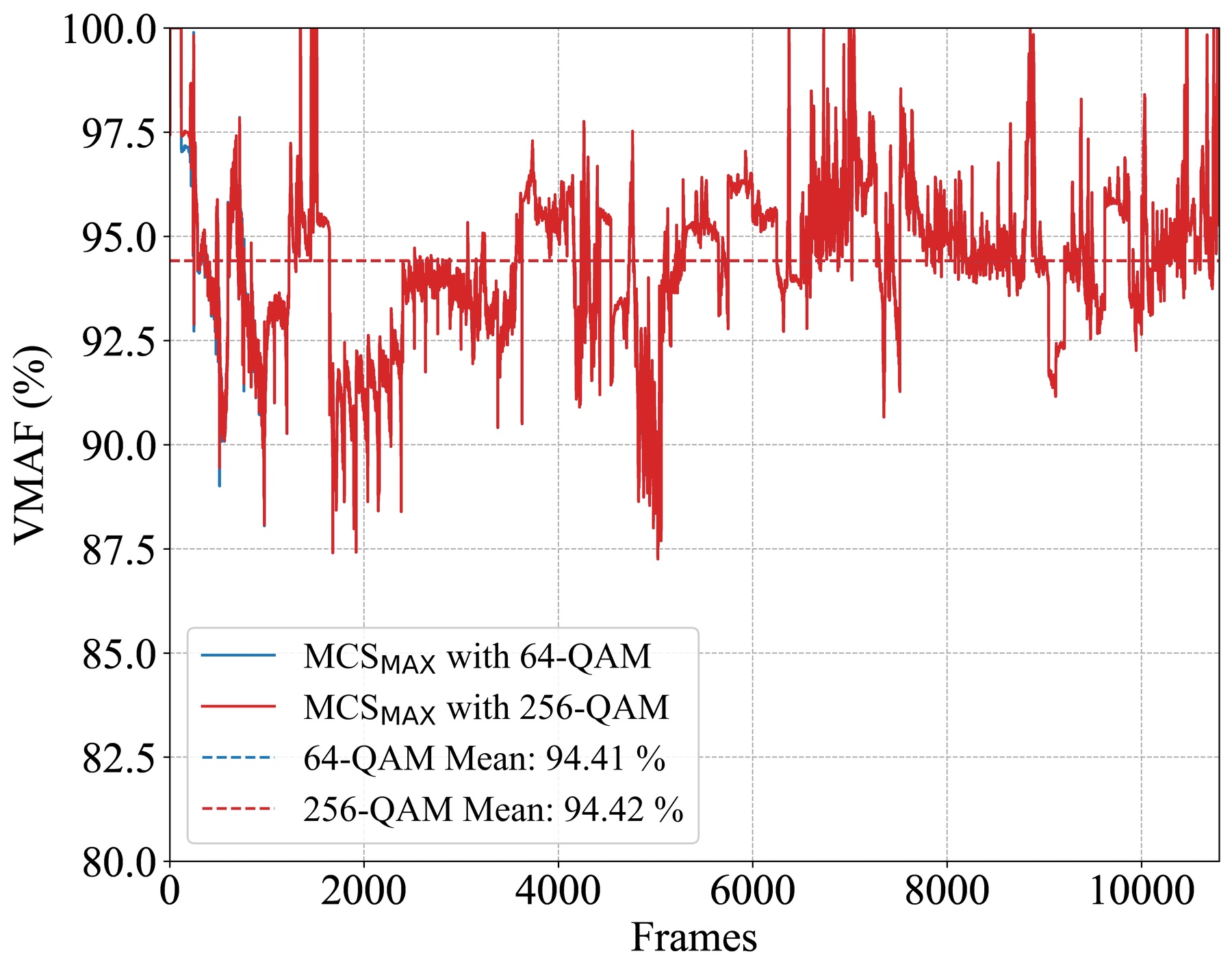}
        \caption{MIMO - 1 User.}
        \label{fig:vmaf_1user_mimo}
    \end{subfigure}
    \begin{subfigure}{0.33\linewidth}
        \centering
        \includegraphics[width=\linewidth]{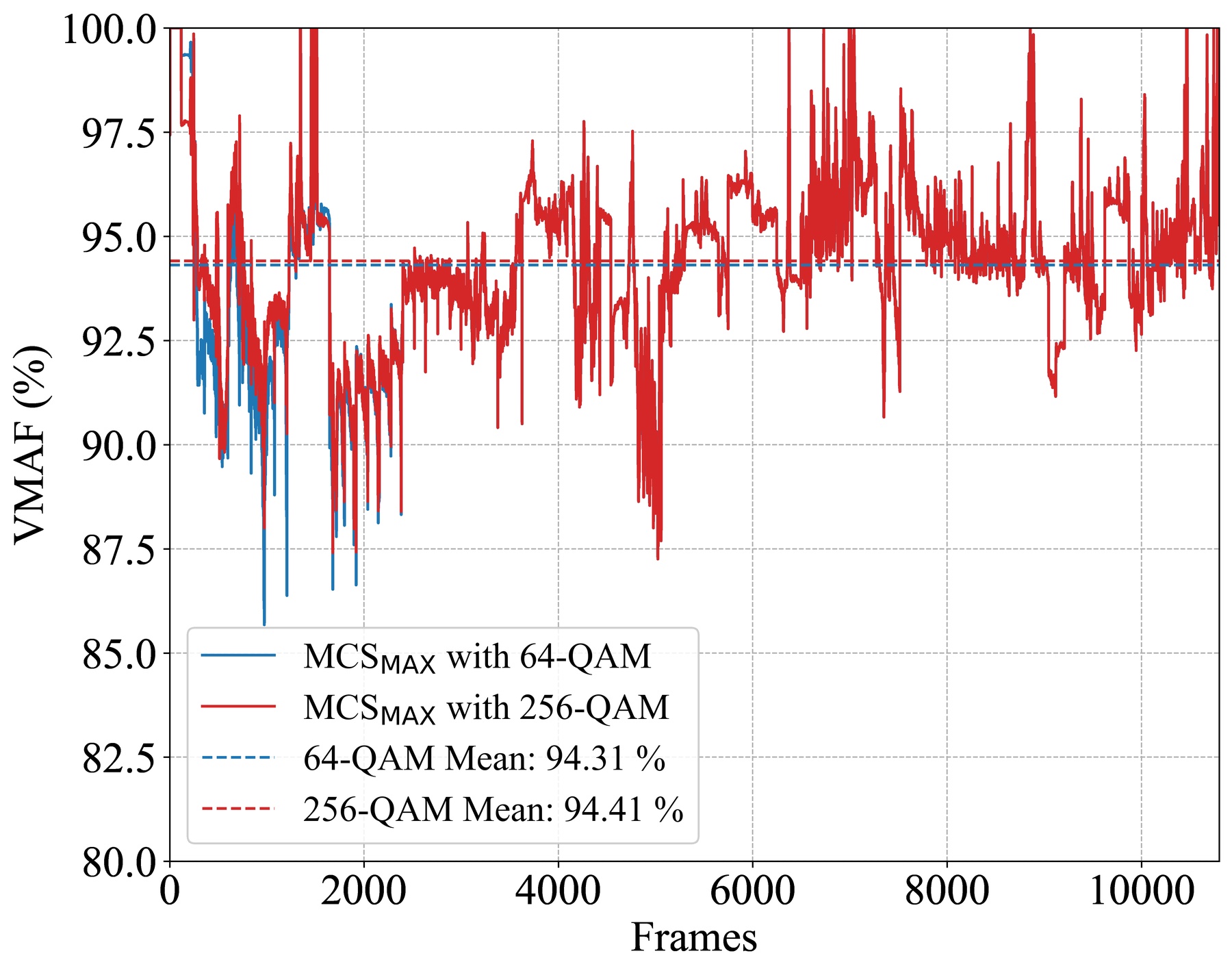}
        \caption{MIMO - 3 Users.}
        \label{fig:vmaf_3user_mimo}
    \end{subfigure}
    \begin{subfigure}{0.33\linewidth}
        \centering
        \includegraphics[width=\linewidth]{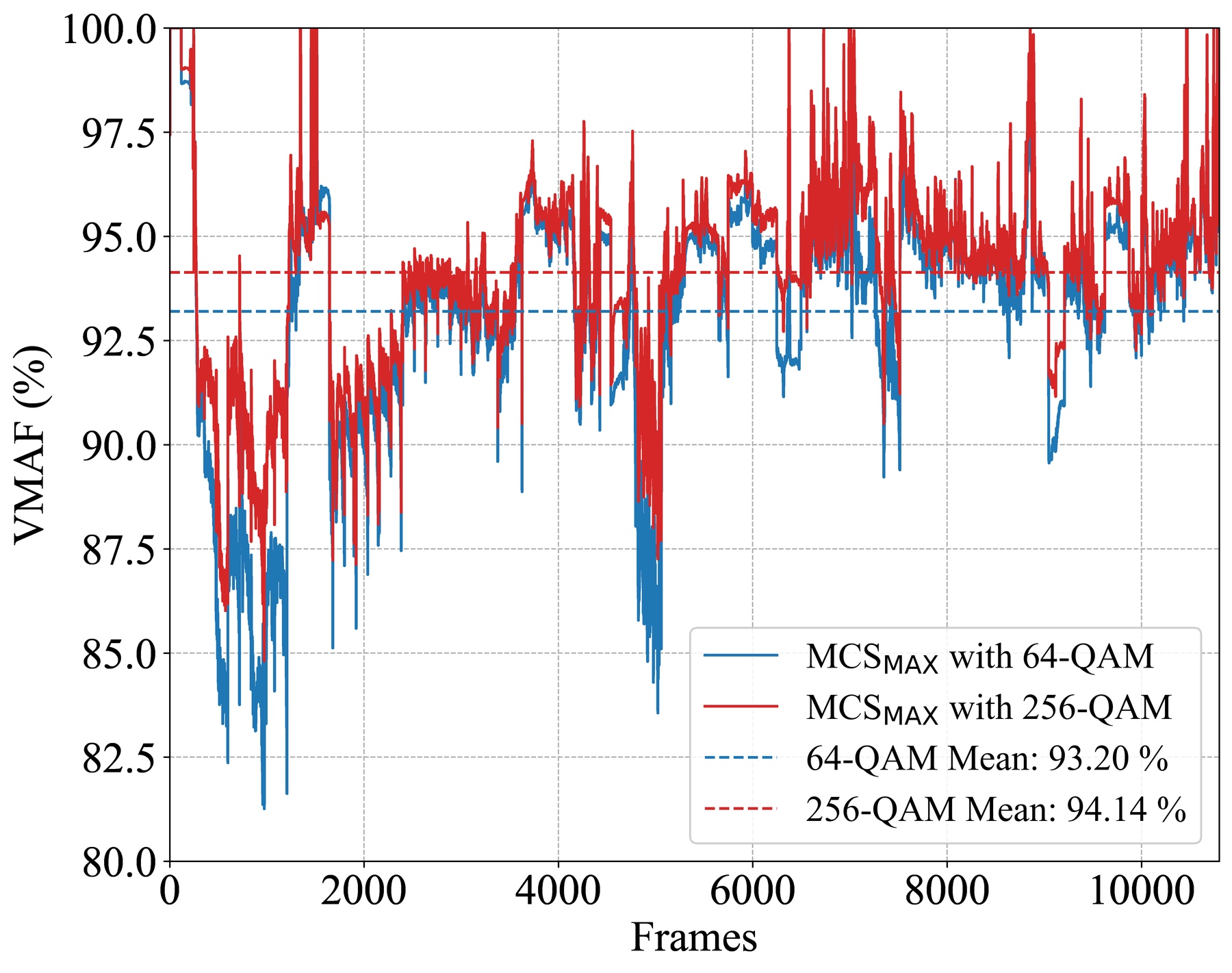}
        \caption{MIMO - 5 Users.}
        \label{fig:vmaf_5user_mimo}
    \end{subfigure}
    \caption{Variation of the VMAF (\%) across the received frames.}
    \label{fig:vmaf}
\end{figure*}

Figure~\ref{fig:vmaf} comes up with the VMAF score (in percentage) computed per OTT video frame across the entire duration of the experiment. According to~\cite{rassool_2017}, a VMAF score of 93\% corresponds to visually indistinguishable quality relative to the original video. As shown in Subfigures~\ref{fig:vmaf_1user_mimo},~\ref{fig:vmaf_3user_mimo}, and~\ref{fig:vmaf_5user_mimo}, this condition is consistently achieved on average across all network load scenarios under the MIMO transmission mode, although occasional frames fall slightly below the threshold. On the contrary, for the SISO configuration, when more than three users share the same FWA link (Subfigures~\ref{fig:vmaf_3user_siso} and~\ref{fig:vmaf_5user_siso}), several frames exhibit VMAF scores below 70\%, indicating noticeable degradation in perceived visual quality.

%According to~\cite{rassool_2017}, a VMAF score of 93 indicates that a video reproduces the original with visually indistinguishable quality. Figs.~\ref{fig:vmaf_1user_mimo},~\ref{fig:vmaf_3user_mimo}, and~\ref{fig:vmaf_5user_mimo} demonstrate that this requirement is met for all network load variations when the MIMO transmission mode is configured. In the SISO configuration, when more than three users are connected to the FWA network (Figs.~\ref{fig:vmaf_3user_siso} and~\ref{fig:vmaf_5user_siso}), frames with VMAF scores less than 70\% occur throughout the video streaming. 

\subsection{General Discussion}

Between frames 1 and 2,200, all three quality indicators exhibit a pronounced degradation under the SISO configuration mode. As shown in Subfigures~\ref{fig:psnr_5user_siso}, ~\ref{fig:ssim_5user_siso} and~\ref{fig:vmaf_5user_siso}, PSNR values experience two distinct drops, falling below 30~dB, while SSIM decreases to below 0.80, suggesting a marked loss of structural fidelity. Correspondingly, VMAF scores drop below 55\%, reflecting perceptible degradation in visual quality according to human perception benchmarks.

Additional performance deterioration occurs between frames 6,000 and 9,000, where PSNR fluctuates between 30 and 35~dB, and SSIM falls to approximately 0.85 or lower. VMAF scores decline to near or below 80\%, revealing a noticeable reduction in perceived quality and underscoring the limitations of the SISO configuration in maintaining consistent video fidelity under load.

\begin{figure}[!ht]
    \centering
    \begin{subfigure}{\linewidth}
        \centering
        \includegraphics[width=\linewidth]{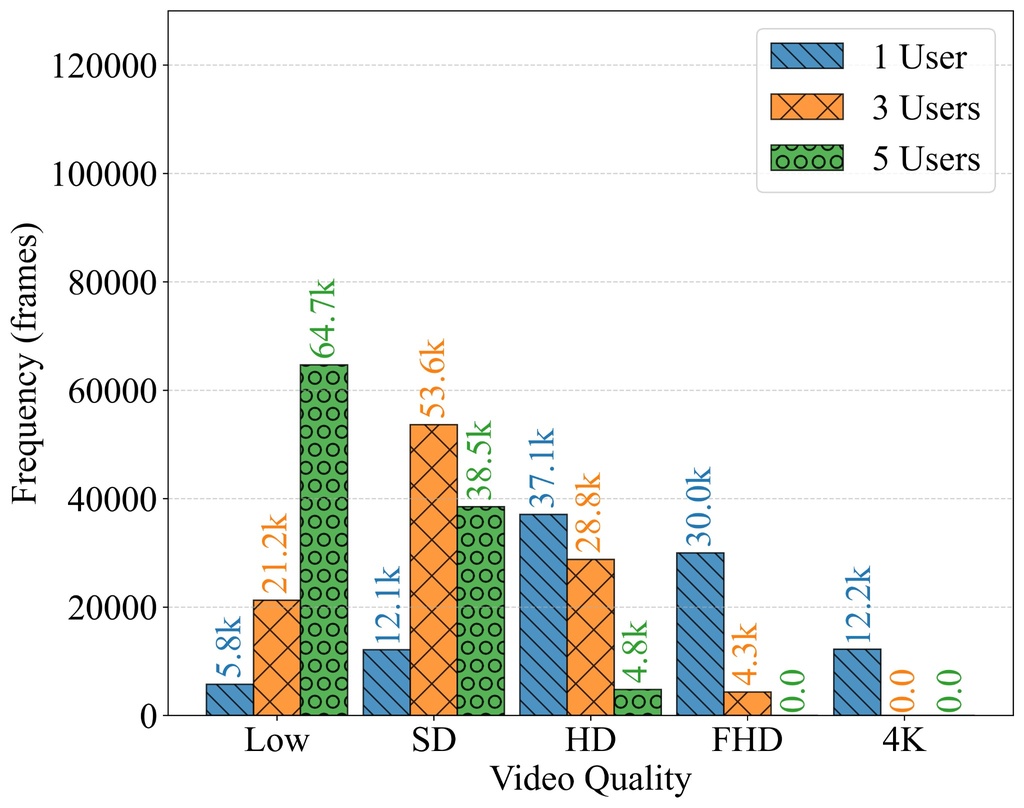}
        \caption{SISO.}
        \label{fig:video_quality_64_siso}
    \end{subfigure}
    \begin{subfigure}{\linewidth}
        \centering
        \includegraphics[width=\linewidth]{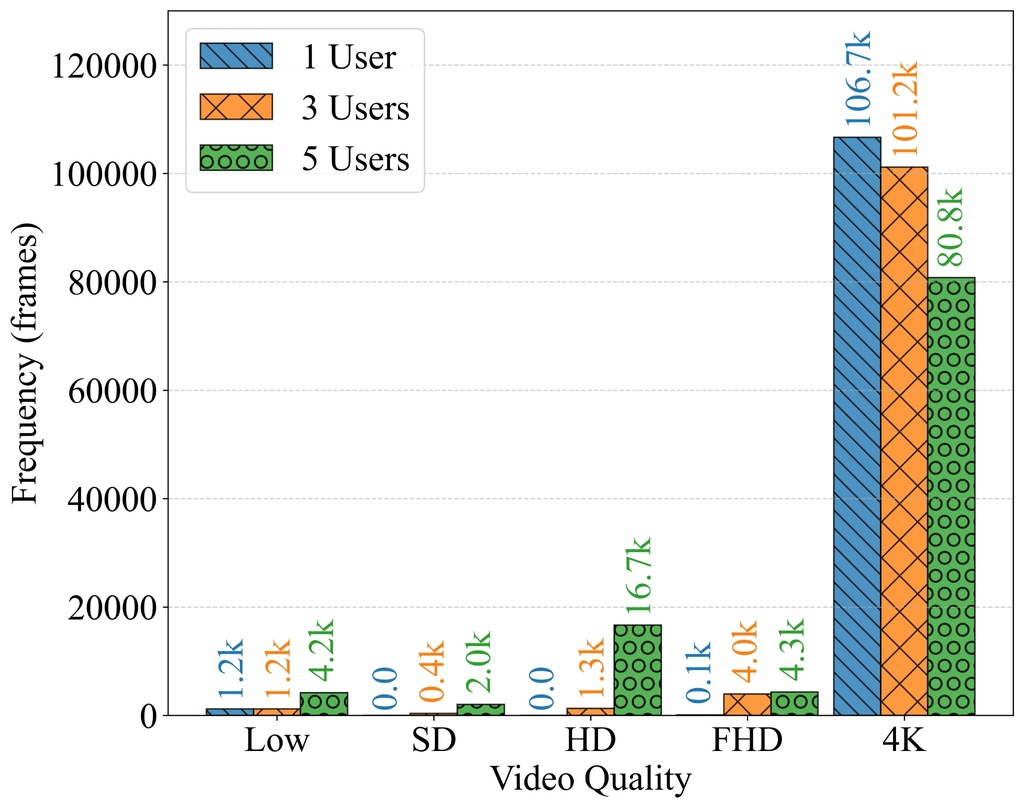}
        \caption{MIMO.}
        \label{fig:video_quality_64_mimo}
    \end{subfigure}
    \caption{Distribution of frame video quality for MCS\textsubscript{max} = 64-QAM.}
    \label{fig:video_quality_64}
\end{figure}

\begin{figure}[!ht]
    \centering
    \begin{subfigure}{\linewidth}
        \centering
        \includegraphics[width=\linewidth]{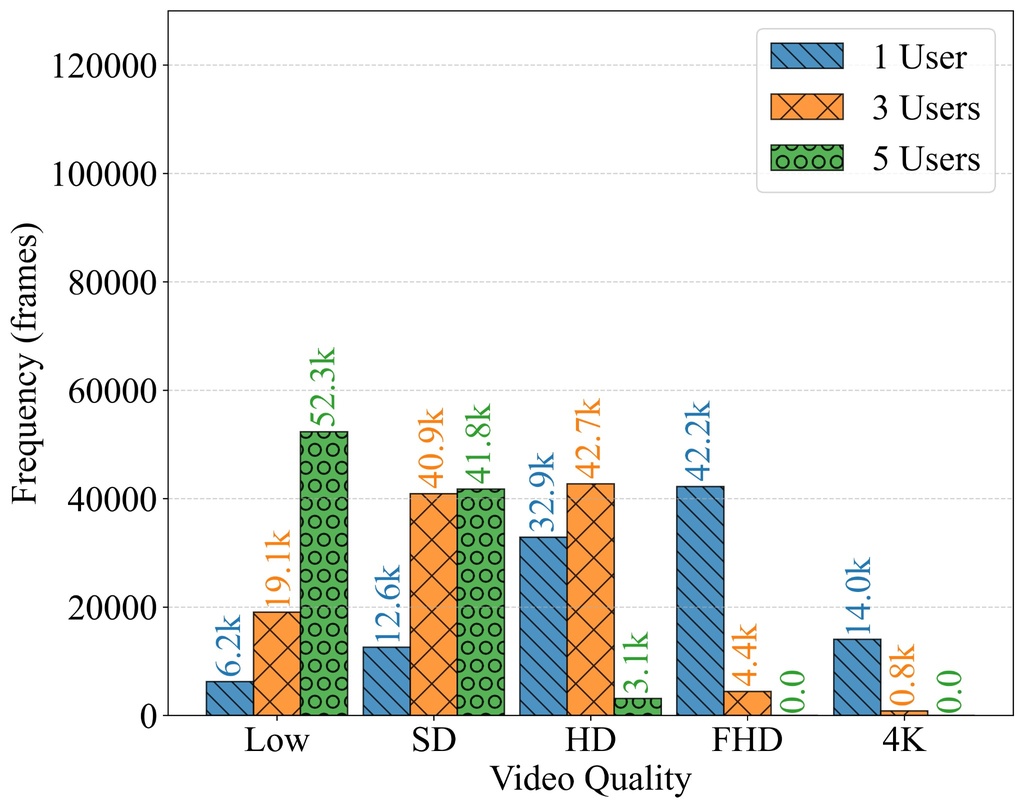}
        \caption{SISO.}
        \label{fig:video_quality_256_siso}
    \end{subfigure}
    \begin{subfigure}{\linewidth}
        \centering
        \includegraphics[width=\linewidth]{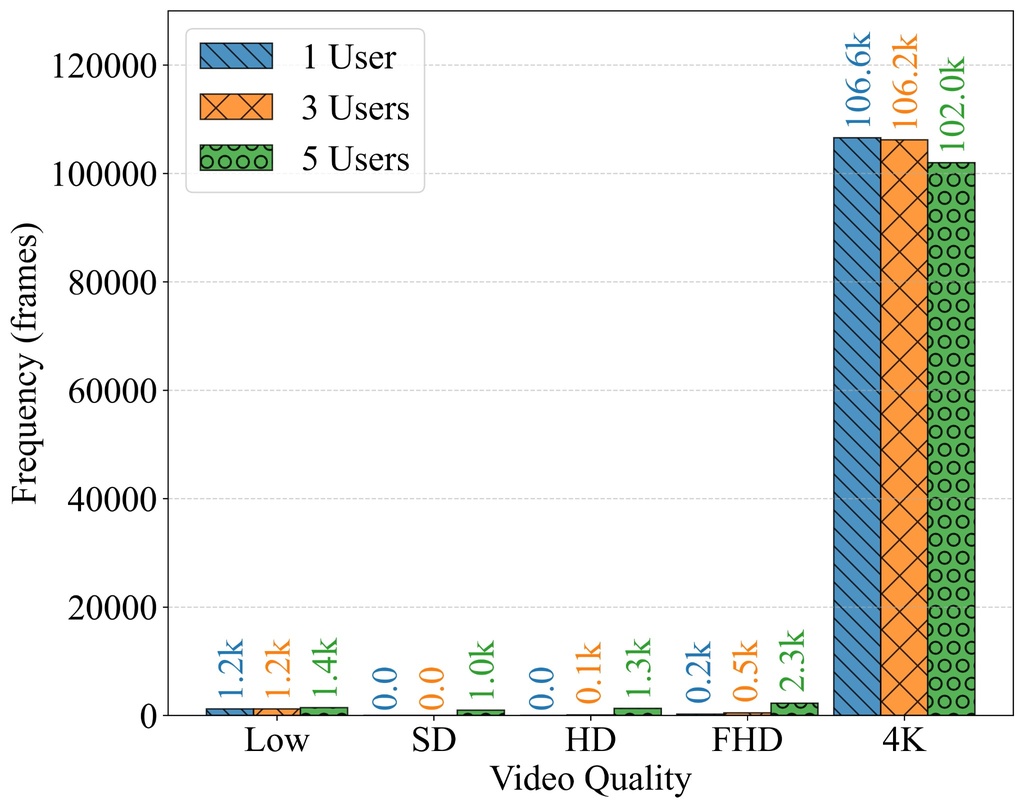}
        \caption{MIMO.}
        \label{fig:video_quality_256_mimo}
    \end{subfigure}
    \caption{Distribution of frame video quality for MCS\textsubscript{max} = 256-QAM.}
    \label{fig:video_quality_256}
\end{figure}

These findings are reinforced by the OTT video quality distribution that Figure~\ref{fig:video_quality_64} illustrates for MCS\textsubscript{max} with 64-QAM and Figure~\ref{fig:video_quality_256} for MCS\textsubscript{max} with 256-QAM. The SISO configuration exhibits a greater concentration of frames in the lower-quality categories for both MCS\textsubscript{max} with 64-QAM (Subfigure~\ref{fig:video_quality_64_siso}) and 256-QAM (Subfigure~\ref{fig:video_quality_256_siso}) modulation schemes, whereas the MIMO configuration maintains a dominant presence in the 4K quality category. 

Furthermore, Figures~\ref{fig:video_quality_64}~and~\ref{fig:video_quality_256} confirm that user-perceived quality decreases as the number of connected users increases. Notably, in the SISO configurations with five users connected (Figure~\ref{fig:video_quality_64_siso}), no frames achieved 4K resolution quality.

%The video quality distribution of the frames, shown in Figure~\ref{fig:video_quality_64} for 64-QAM and Figure~\ref{fig:video_quality_256} for 256-QAM, confirms this behavior. SISO configuration had a higher concentration of frames in the lower quality categories for both 64-QAM (Subfigures~\ref{fig:video_quality_64_siso}) and 256-QAM (Fig.~\ref{fig:video_quality_256_siso}) configurations. In contrast, the MIMO configuration had the highest concentration in the 4K quality category. Figures~\ref{fig:video_quality_64} and~\ref{fig:video_quality_256} also confirm that the user experience decreases as the number of connected users increases, particularly in the 4K category, when SISO was configured and five users were connected. 

In general, while the SISO configuration with MCS\textsubscript{max} set to 64-QAM may fall short of meeting the performance requirements for OTT applications under high network load, it represents a practical trade-off at lower loads and in MIMO scenarios, where the use of simpler modulation can deliver meaningful energy savings on both gNB and UE sides.

\section{Conclusion and Future Work}

This paper puts forward a comprehensive experimental assessment of the video quality of a UHD OTT application service over a network using 5G FWA connectivity, analyzing the effects of variations in transmission mode, modulation scheme, and network load. By combining a real-world 5G telco-cloud testbed built on open-source components with objective full-reference metrics (PSNR, SSIM, and VMAF), we quantified the relationship between network performance (QoS) and perceived visual quality (QoE) in representative FWA scenarios. 

The results suggest that the MIMO configuration consistently outperforms SISO, maintaining the OTT application content with higher throughput and superior video quality across all network load levels. In particular, the combination of MIMO and MCS\textsubscript{max} with 256-QAM sustains VMAF scores above the visually lossless threshold, even with five concurrent users, whereas SISO configurations exhibit noticeable degradation under increased load. Moreover, the influence of modulation order is proven to be more significant in MIMO setups, while its impact in SISO remains marginal. These findings highlight that spatial diversity and higher-order modulation are essential for affording high-quality OTT UHD content delivery in 5G FWA-based scenarios. The adopted methodology that integrates automated campaigns, DASH-based adaptive streaming, and frame-level quality analysis demonstrated effectiveness for linking network-level QoS parameters to QoE.

While the empirical results presented in this study were obtained using a specific 5G telco-cloud testbed and spectrum configuration, the underlying assessment methodology, which integrates automated application-layer segment acquisition with frame-level objective metrics, is inherently platform-agnostic. By leveraging open-source components and standard MPEG-DASH protocols, this framework provides a reproducible baseline that can be generalized to other 5G FWA deployments, different frequency bands, or large-scale commercial network infrastructures.

%Future work will explore additional adaptive bitrate algorithms, extended bandwidth allocations, RAN and 5GC disaggregation configurations, as well as experiments under dynamic channel and mobility conditions. These extensions aim to advance the optimization of FWA deployments for next-generation applications that demand immersive, low-latency, and bandwidth-intensive content delivery.

To further strengthen the contextual grounding and improve the generalization of these findings, future work will involve exploring RAN and 5GC disaggregation configurations to evaluate how architectural shifts impact UHD content delivery. Additionally, we plan to extend our experimental campaigns to include dynamic channel conditions and mobility scenarios, moving beyond the current static testbed to assess the robustness of 5G FWA infrastructures across a broader range of deployment environments. This roadmap includes evaluating next-generation applications, such as immersive and high-motion content, which are significantly more bandwidth-intensive and require lower latency. By testing these diverse content profiles under varying network loads, we aim to provide comprehensive deployment strategies for high-quality video services that require both high bandwidth and high reliability in complex operational environments.

\label{sec:conclusions}

% Referências a partir do arquivo referencias.bib
\bibliographystyle{IEEEtran}
\bibliography{referencias}

% Introdução 2
% II Métricas de avaliação 2
% III Metodologia dos Testes 3
% III-.1 Configuração do ambiente
% do servidor . . . . . . . . . . 3
% III-.2 Codificação de áudio e vídeo 3
% III-.3 Geração do Manifesto DASH 4
% III-.4 Captura dos pacotes . . . . . 4
% III-.5 Reconstrução de vídeos . . . 4
% III-.6 Cálculo de Métricas . . . . . 4
% IV Resultados 4
% V Conclusões 

% Nelson
\begin{IEEEbiography}
    [{\includegraphics[width=1in,height=1.25in,clip,keepaspectratio]{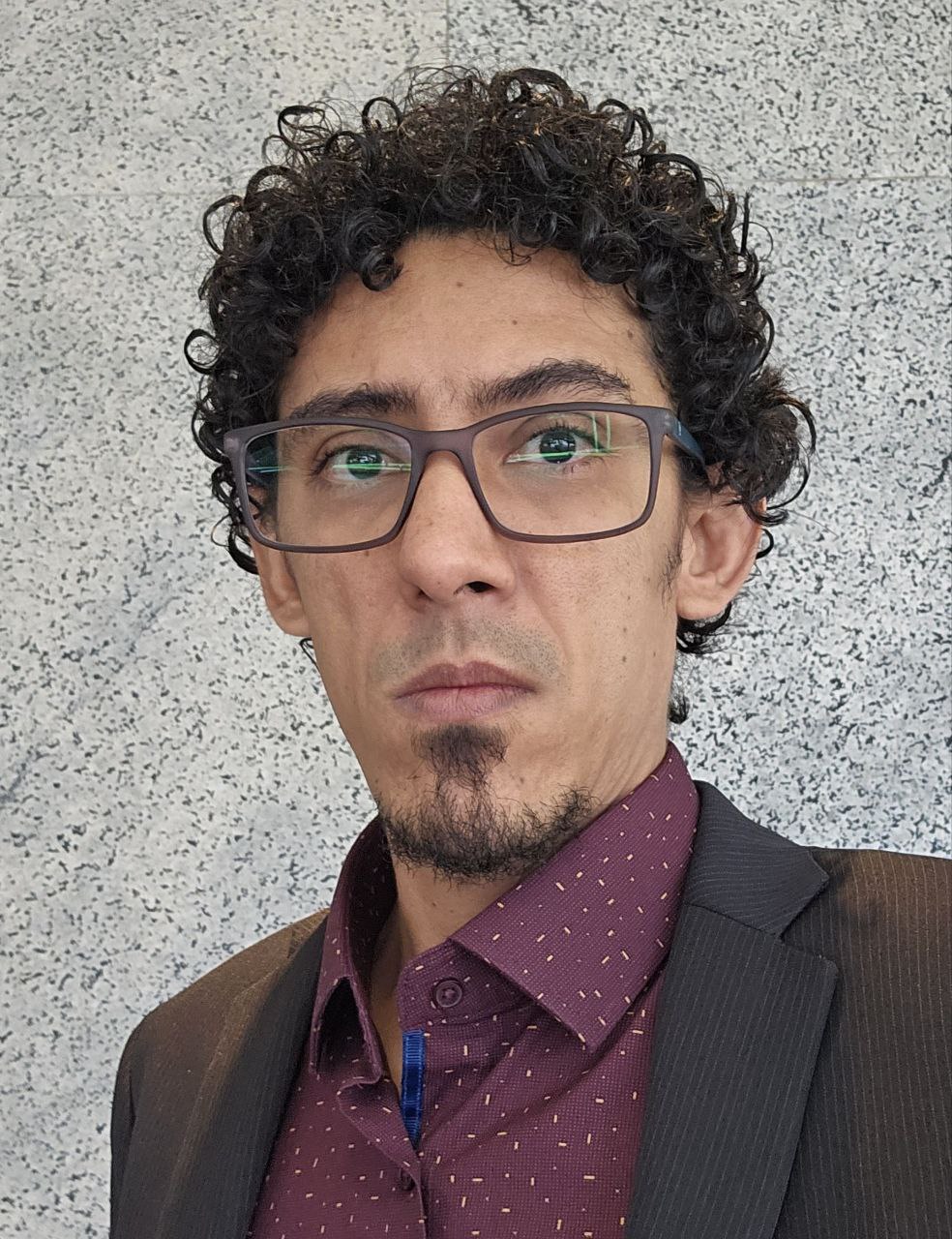}}]{Nelson Ion de Oliveira}
    is a Professor at the Digital Metropolis Institute (IMD) of the Federal University of Rio Grande do Norte (UFRN). He obtained a degree in Technology in Systems Analysis and Development (2011) from the Federal Institute of Education, Science and Technology of Rio Grande do Norte (IFRN), received his M.Sc. in Systems and Computing from UFRN in 2017, and is currently a Ph.D. candidate in Systems and Computing at UFRN. From 2012 to 2014, he was a Temporary Professor at IFRN. From 2015 to 2018, he served as an Assistant Professor at IMD/UFRN and has been an Adjunct Professor in the same department since 2018. He is the author of three didactic books for technical (high school) courses. Oliveira has worked with Open RAN solutions for Lenovo and is a member of the Leading Advanced Technologies Center of Excellence (LANCE).
\end{IEEEbiography}

% Flavio
\begin{IEEEbiography}
[{\includegraphics[width=1in,height=1.25in,clip,keepaspectratio]{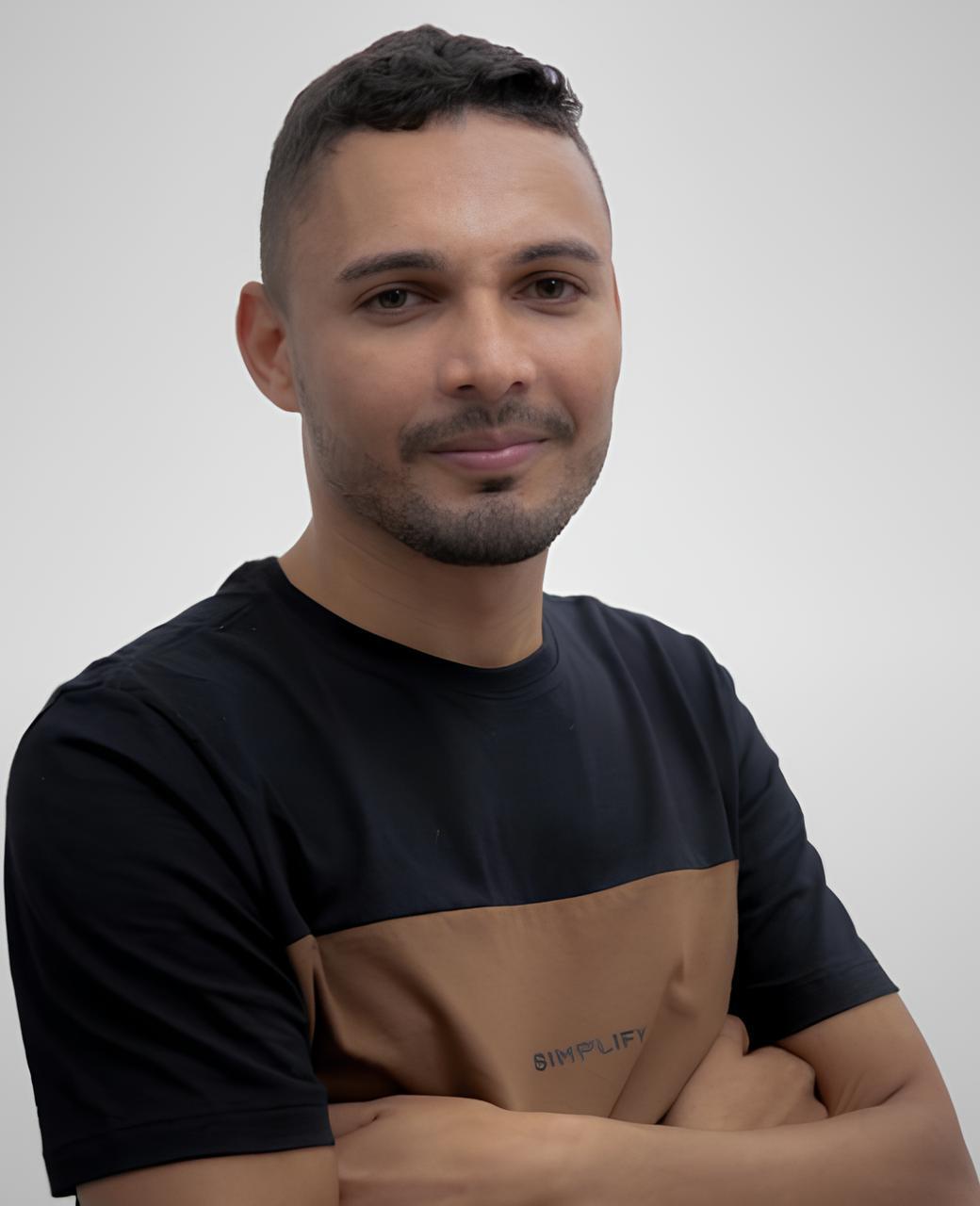}}]
{Flavio de Andrade Silva} holds a Bachelor’s degree in Science and Technology from UFRN (Federal University of Rio Grande do Norte), earned in 2021. He is currently pursuing a Bachelor’s degree in Telecommunications Engineering at the same institution, and works as a DevOps engineer. He has experience in the field of computer networks and with technologies such as microservices, automation, CI/CD pipelines, observability, and cloud-native solutions.
He was a research fellow in the project “Self-Organized Management of Network Slices over Multi-RAN Network Infrastructures”, conducted at the Leading Advanced Technology Center of Excellence (LANCE), where he contributed to OpenRAN network deployment projects using OpenAirInterface (OAI) and Open5GS, as well as to the deployment of the Magma platform for LTE, 5G, and Wi-Fi networks.
\end{IEEEbiography}

% Paulo
\begin{IEEEbiography}
[{\includegraphics[width=1in,height=1.25in,clip,keepaspectratio]{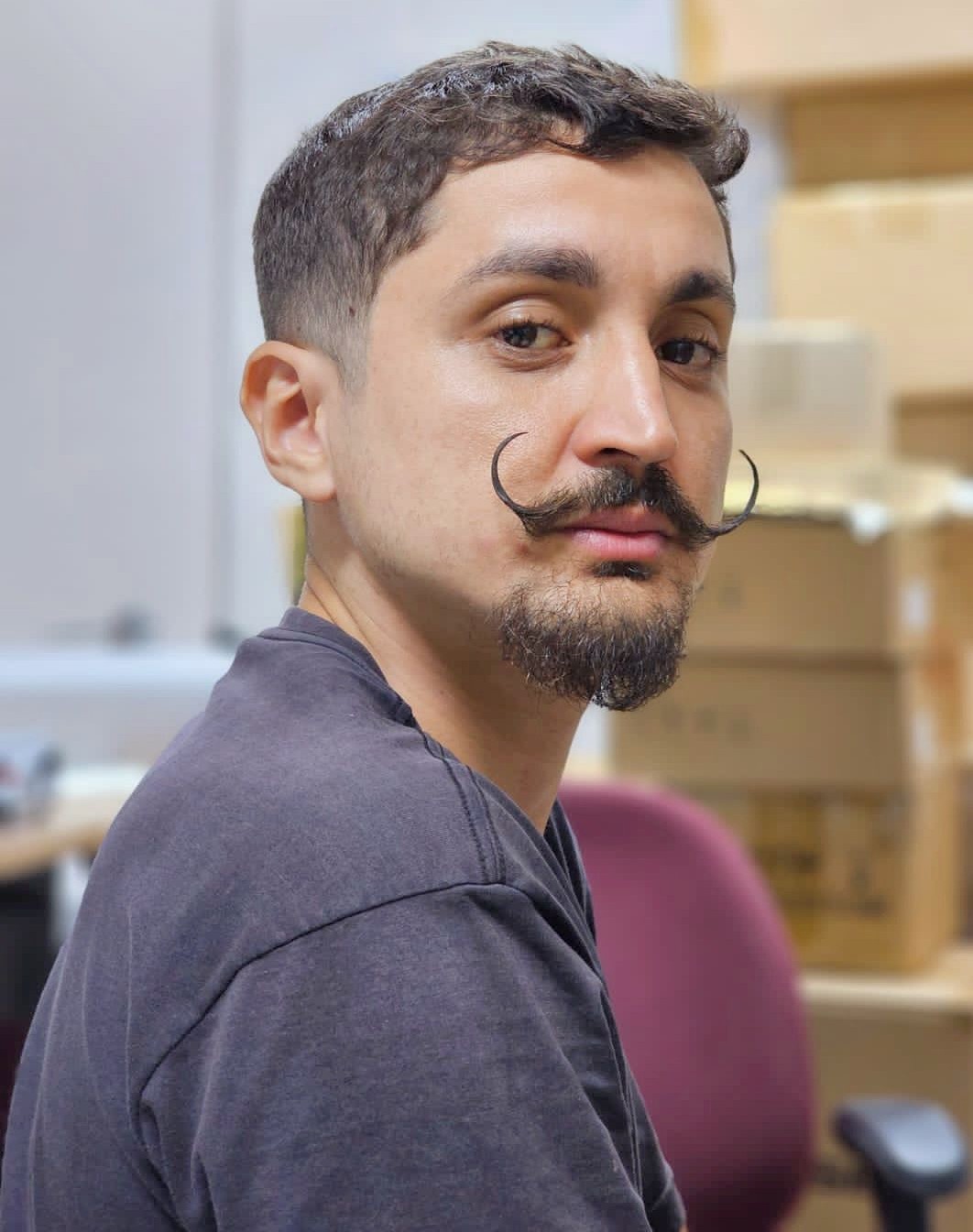}}]
{Paulo Eduardo da Silva Junior} He holds a technical degree in Computer Science from IFRN (2013) and earned his Bachelors degrees in Science and Technology (2017) and Telecommunications Engineering (2022) from the Federal University of Rio Grande do Norte (UFRN). He is currently pursuing a Masters degree in Electrical and Computer Engineering at UFRN, focusing on E2SMs and KPI monitoring through the E2 interface in Open RAN systems. He is a researcher at the Center of Excellence in Advanced Technologies (LANCE) and a member of the REGINA Research Group on Future Internet Services and Applications. From 2023 to 2024, he contributed to the deployment and structuring of the LANCE Laboratory as part of the Lenovo Center of Excellence in Advanced Networks, supporting the development of a complete 5G infrastructure for experimentation, validation, and demonstration of innovative Open RAN use cases. Paulo has solid experience in monitoring distributed systems and building observability dashboards, as well as in container-based environments, automation pipelines, CI/CD workflows, and cloud-native solutions. He also works with Open RAN implementations using platforms such as OpenAirInterface (OAI), Open5GS, srsRAN, FlexRIC, and the SC RIC.
\end{IEEEbiography}

% Ricardo
\begin{IEEEbiography}
[{\includegraphics[width=1in,height=1.25in,clip,keepaspectratio]{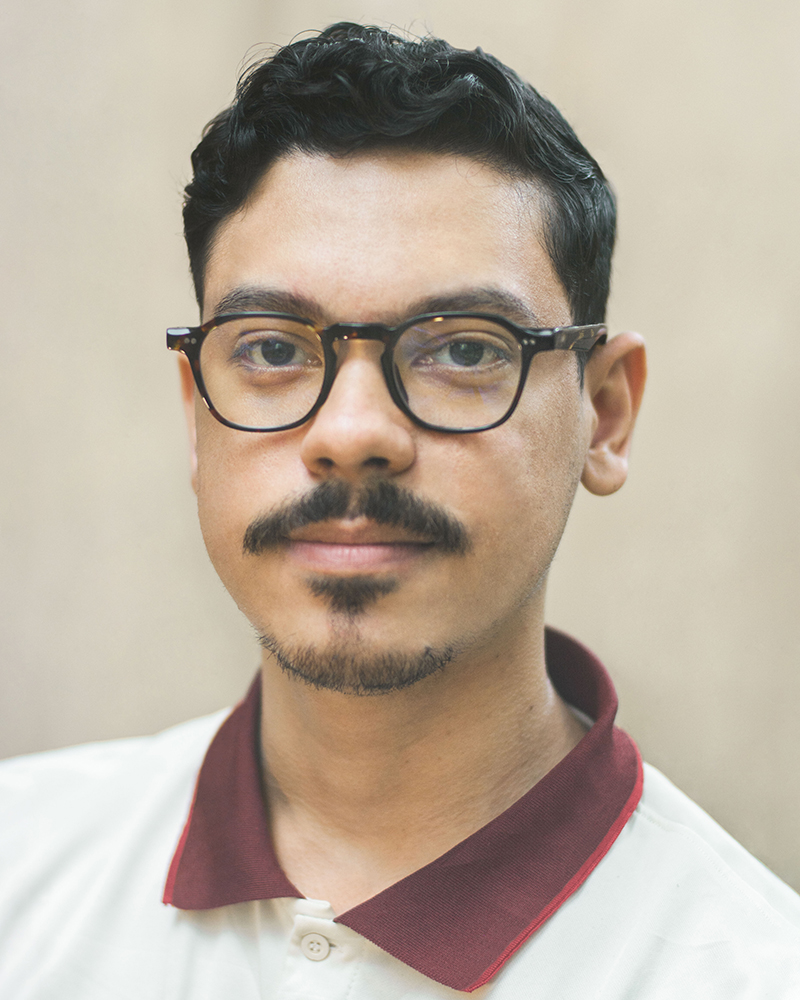}}]
{Ricardo Q. de F. H. Silva} holds a B.S. (2023) and an M.Sc. (2025) in Electrical Engineering from the Federal University of Rio Grande do Norte (UFRN), Brazil. He also earned a technical degree in Electronics from the Federal Institute of Education, Science, and Technology of Rio Grande do Norte (IFRN) in 2017. Currently, he is a Ph.D. candidate and research fellow at the GppCom Research Group, where his work centers on mobile communication systems. Silva has solid experience with network simulation using the ns-3 discrete-event simulator, particularly in Wi-Fi, LTE, and 5G NR scenarios. He also works on OpenRAN deployments leveraging the OpenAirInterface (OAI), open5GS, and srsRAN platforms. Additionally, he contributes to projects focused on RF-EMF exposure assessment, in collaboration with the Brazilian National Telecommunications Agency (ANATEL).
\end{IEEEbiography}

% Mathews
\begin{IEEEbiography}
[{\includegraphics[width=1in,height=1.25in,clip,keepaspectratio]{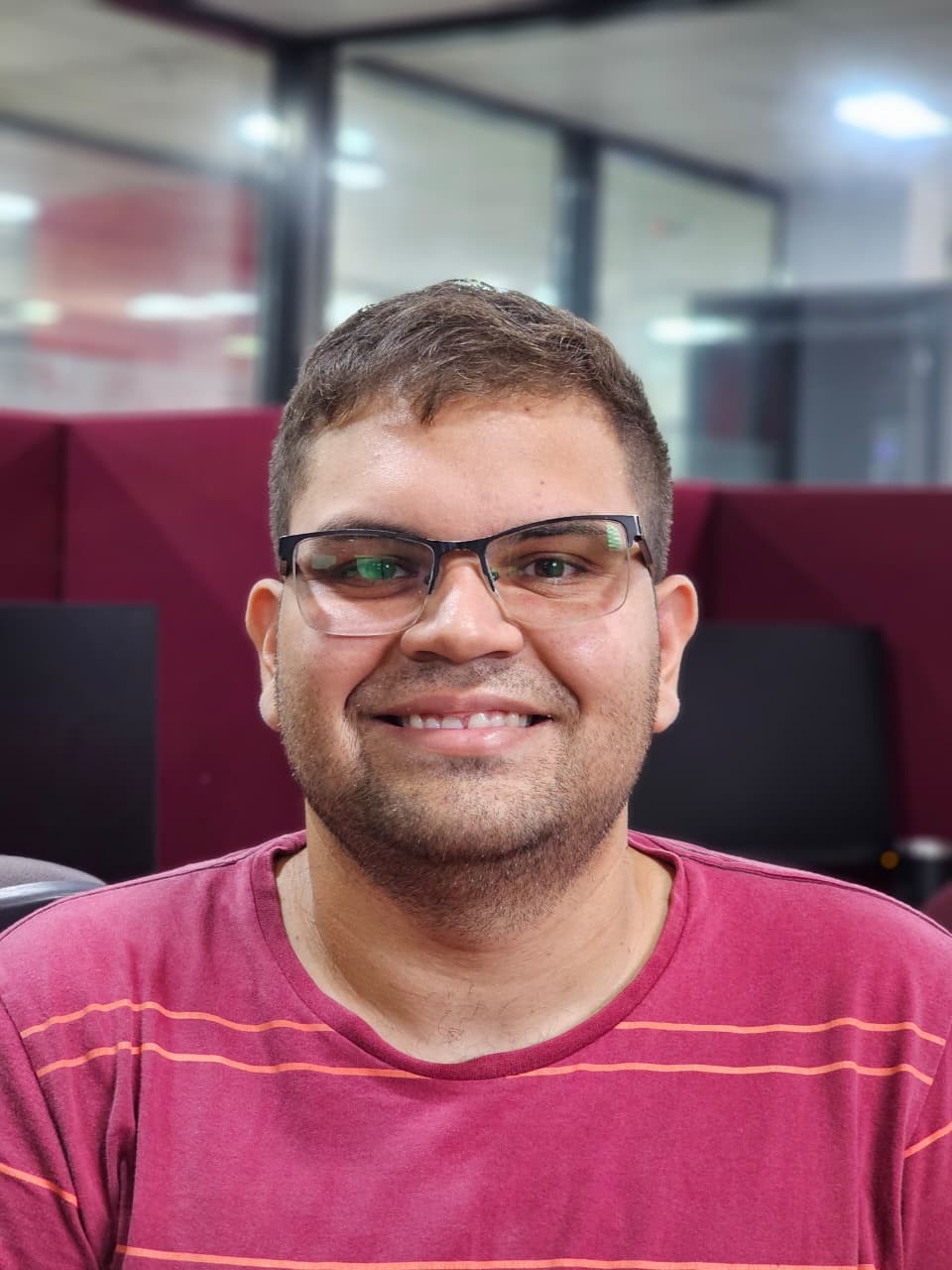}}]
{Mathews P. S. Lima} holds a degree in Computer Networks Technology (2018) from the Federal Institute of Education, Science and Technology of Rio Grande do Norte (IFRN) and an M.Sc. in Systems and Computing from the Federal University of Rio Grande do Norte (UFRN), obtained in 2021. Currently pursuing his Ph.D. in Computer Science at UFRN. He is a researcher at the Leading Advanced Technologies Center of Excellence (LANCE) and a member of the Research Group in Future Internet Service and Applications (REGINA). His research interests include 5G, mobility management, network slicing and machine learning.
\end{IEEEbiography}

% Daniel
\begin{IEEEbiography}
[{\includegraphics[width=1in,height=1.25in,clip,keepaspectratio]{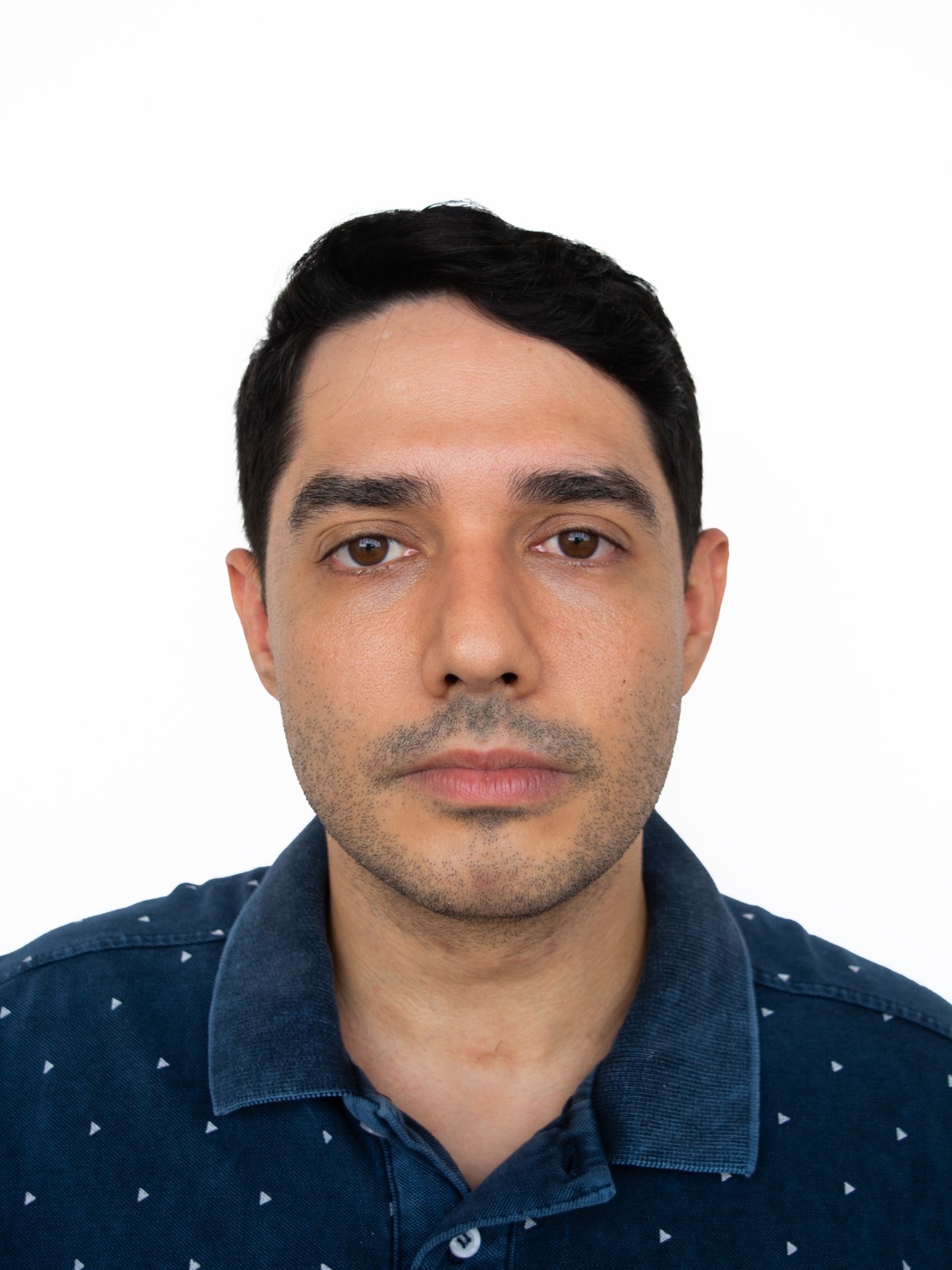}}]
{Daniel Rodrigues de Luna} holds a B.Sc. in Science and Technology (2013) and a B.Eng. in Telecommunications Engineering (2016), followed by an M.Sc. (2018) and Ph.D. (2023) in Electrical and Computer Engineering with a focus on Telecommunications Systems (Mobile Communications), all from the Federal University of Rio Grande do Norte (UFRN). His main area of expertise is prototyping computational tools for Radio Resource Management (RRM) algorithms and emulating 5G/6G network cores/RAN via O-RAN using USRPs. Also, develops and evaluates Wi-Fi, LTE, NR, and IoT systems (UAN, NB-IoT, LoRa) via open-source platforms like ns-3. Research interests integrate AI with telecom systems, emphasizing reinforcement learning for dynamic IoT networks and AI-driven E2E slicing in Open RAN (O-RAN/OAI). Currently, He is post-doctoral researcher focusing on ubiquitous IoT connectivity through private 5G/Wi-Fi converged networks.
\end{IEEEbiography}

% Antonio
\begin{IEEEbiography}[{\includegraphics[width=1in,height=1.25in,clip,keepaspectratio]{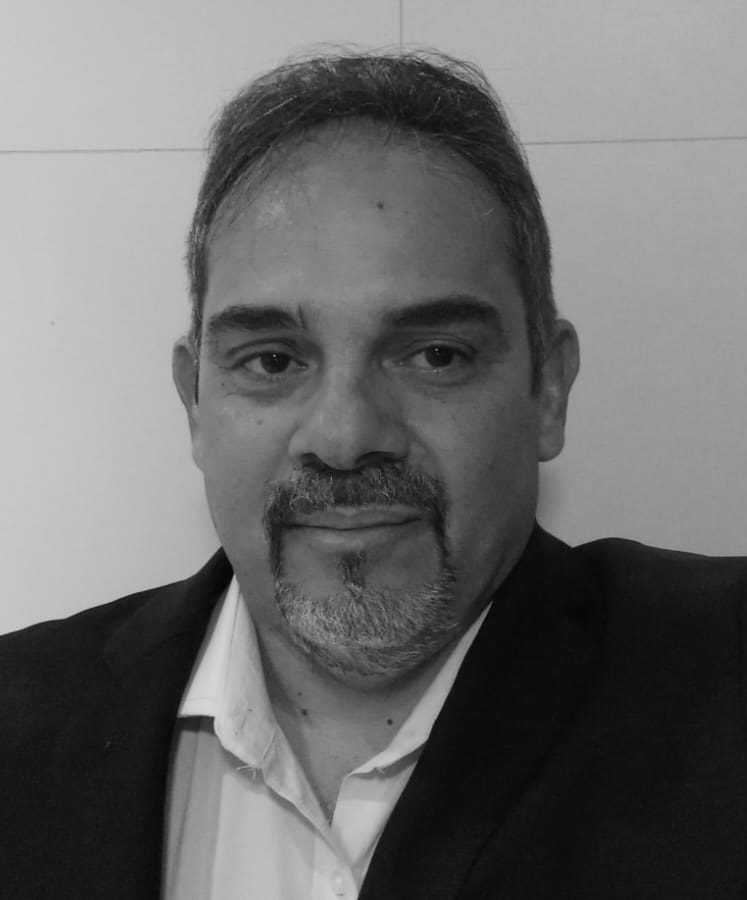}}]{Antonio Luiz Pereira de Siqueira Campos}
received the B.Sc. and M.Sc. degrees in Electrical Engineering from the Federal University of Rio Grande do Norte (UFRN) in 1996 and 1999, respectively, and the Ph.D. degree in Electrical Engineering from the Federal University of Paraíba (UFPB) in 2002. He is currently an Associate Professor IV at the Federal University of Rio Grande do Norte, where he teaches undergraduate courses in Telecommunications Engineering. He also serves as a faculty member in the Graduate Program in Electrical and Computer Engineering (PPgEEC/UFRN), supervising M.Sc. and Ph.D. students and teaching courses in the field of Telecommunications. Prof. Antonio Campos is the leader of the research group "Study of Periodic and Quasi-Periodic Structures and their Applications in Telecommunications" at UFRN. His research interests include electromagnetic theory, microwaves, and instrumentation applied to telecommunications, with particular focus on periodic structures, numerical and computational methods, microstrip antennas and filters, and telecommunication systems. He has authored over 200 scientific publications, including books, book chapters, journal articles, and conference papers. He has been an active member of the Brazilian Society of Microwaves and Optoelectronics (SBMO) since 2005.
\end{IEEEbiography}

% Augusto
\begin{IEEEbiography}
[{\includegraphics[width=1in,height=1.25in,clip,keepaspectratio]{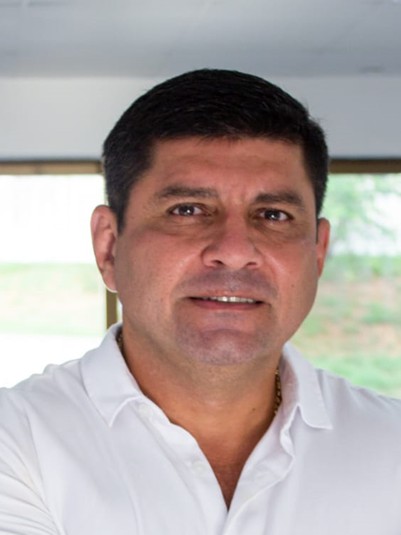}}]
{Augusto Venâncio Neto} (IEEE Senior Member) is a Full Professor at the Department of Informatics and Applied Mathematics (DIMAp) of the Federal University of Rio Grande do Norte (UFRN), Natal/RN, Brazil; Productivity fellow at the National Council for Scientific and Technological Development (CNPq); and member of the Instituto de Telecomunicações, Portugal. He is the leader of the Research Group in Future Internet Service and Applications (REGINA), with background in cutting-edge technologies in computer networks and telecommunications, through which managing large research and developing teams, allowed him to form tens of PhDs and masters’ researchers, as well as for authoring/coauthoring more than 250 papers, including patents and several best-paper awards. He is involved in the organization of international flagship conferences and workshops, as well as serving as Guest Editor for special issues of peer-reviewed scholarly journals. His research interests fall in the fields of 5G/6G Networks, AI-supported networking, Mobile Computing, Smart Spaces, IoT, SDN, NFV and Cloud Computing.
\end{IEEEbiography}

% Vicente
\begin{IEEEbiography}
[{\includegraphics[width=1in,height=1.25in,clip,keepaspectratio]{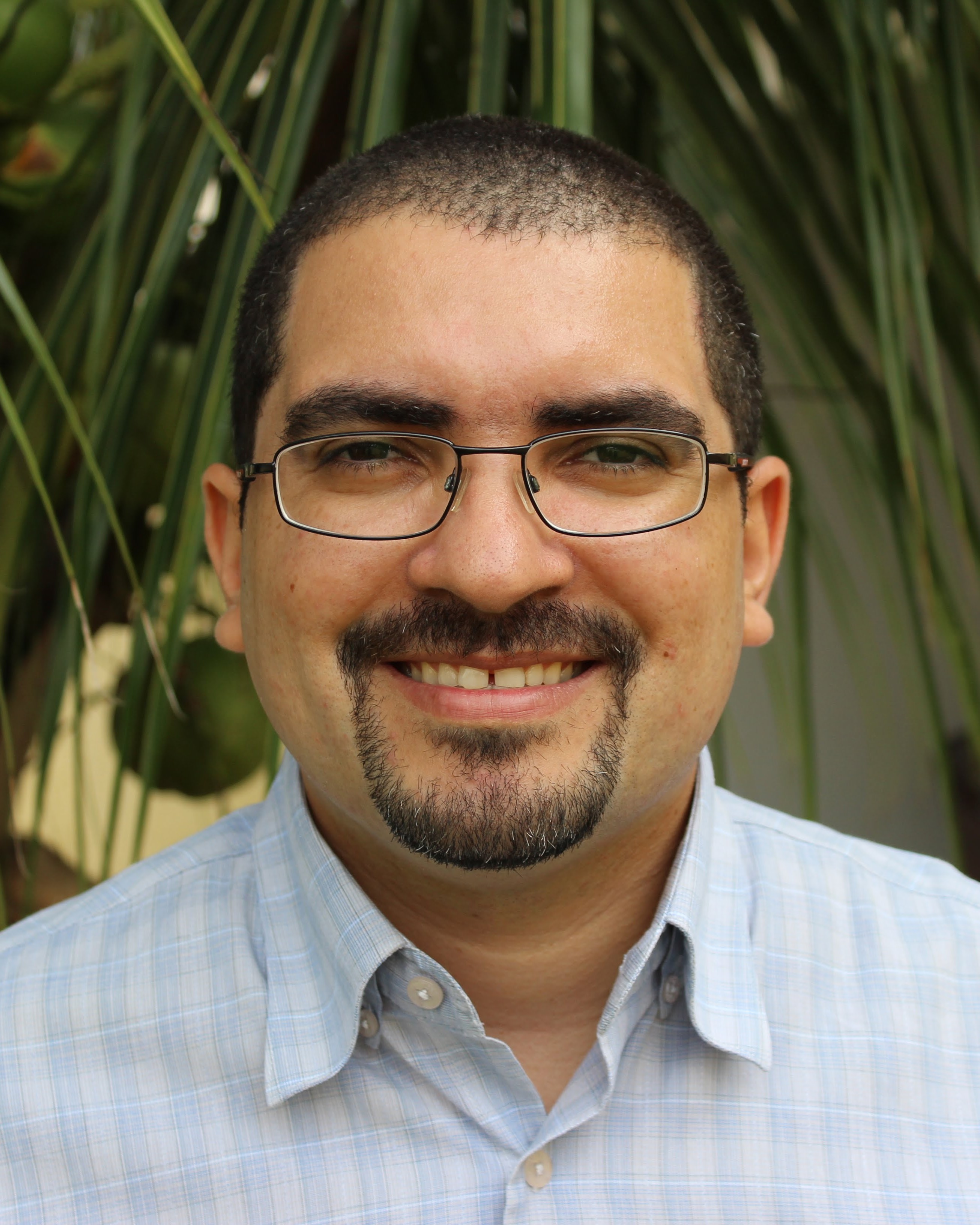}}]
{Vicente A. de Sousa Jr.} received his B.S., M.S and Ph.D.  degrees in Electrical Engineering from the Federal University of Ceará (UFC), Fortaleza, CE, Brazil, in 2001,  2002  and  2009,  respectively.  Between 2001 and 2006, he developed solutions to UMTS/WLAN interworking for   UFC   and   Ericsson of   Brazil. Between 2006 and 2010, he contributed to WIMAX standardization and Nokia’s product as a researcher at the Institute of Technological Development (INdT). Dr.  Sousa is now a professor and the head of the GppCom Research Group at UFRN. He contributed to 5G open RAN projects supported by Lenovo and to NIR measurements and is working with evaluation projects supported by ANATEL Agency.
\end{IEEEbiography}

\EOD

\end{document}